\documentclass{aastex}

\usepackage{emulateapj5,apjfonts}

\input{epsf}

\newcommand\cminvsq{\,{\rm cm}^{-2}}

\newcommand\counts{\,{\rm counts}}

\newcommand\ph{\,{\rm ph}}
\newcommand\secinv{\,{\rm s}^{-1}}

\newcommand\rmo{{\rm o}}

\newcommand{\grad}{{\rm o}}

\newcommand{\bez}{\begin{eqnarray*}}
\newcommand{\eez}{\end{eqnarray*}}
\newcommand{\be}{\begin{equation}}
\newcommand{\ee}{\end{equation}}
\newcommand{\beq}{\begin{eqnarray}}
\newcommand{\eeq}{\end{eqnarray}}
\newcommand{\bc}{\begin{center}}
\newcommand{\ec}{\end{center}}

\begin{document}

\slugcomment{The Astrophysical Journal, in press}

\shortauthors{STERN ET AL.}
\shorttitle{OFF-LINE SCAN OF THE BATSE DAILY RECORDS}

\title{
An Off-line Scan of the BATSE Daily Records and a Large Uniform Sample of 
Gamma-Ray Bursts
}

\author{Boris E. Stern,\altaffilmark{1,2,3} 
Yana Tikhomirova,\altaffilmark{2,3} 
Dmitrii Kompaneets,\altaffilmark{2} 
Roland Svensson,\altaffilmark{3} and
Juri Poutanen\altaffilmark{3}} 

\altaffiltext{1}{Institute for Nuclear Research, Russian Academy of Sciences,
Moscow 117312, Russia}

\altaffiltext{2}{Astro Space Center of Lebedev Physical Institute,
 Profsoyuznaya 84/32, Moscow 117810, Russia; 
 stern,dkompan@lukash.asc.rssi.ru,
jana@anubis.asc.rssi.ru}

\altaffiltext{3}{Stockholm Observatory, SE-133 36 Saltsj\"obaden, Sweden;
stern, jana, svensson, juri@astro.su.se}

\begin{abstract}  
During a scan of the archival  BATSE daily records  covering the entire 9.1
years (TJD  8369--11690)  of the BATSE  operation,  3906  gamma-ray  bursts
(GRBs) have been detected.  2068 of these GRBs are  previously  known BATSE
triggers while 1838 of them are new non-triggered  bursts.  It is important
that all events were  detected in the same type of data and were  processed
with the same  procedure.  Therefore  these 3906 GRBs  constitute a uniform
sample.  We  have   created  a   publically   available   electronic   data
base\footnotemark[4] containing this sample.  We describe the procedures of
the  data  reduction,  the  selection  of  the  GRB  candidates,   and  the
statistical tests for possible non-GRB  contaminations.  We also describe a
novel  test  burst  method  used to  measure  the scan  efficiency  and the
information  obtained using the test bursts.  Our scan  decreases the BATSE
detection threshold to $\sim 0.1 \ph \secinv \cminvsq$.  As a first result,
we show that the  differential  $\log N-\log P$ distribution  corrected for
the detection efficiency extends to low brightnesses without any indication
of a turn-over.  Any reasonable  extrapolation of the new $\log N - \log P$
to lower  brightnesses  imply a rate of  several  thousands  of GRBs in the
Universe per year. 
\end{abstract}

\section{Introduction} 

\footnotetext[4]{http://www.astro.su.se/groups/head/grb\_archive.html}
 
The sample of gamma-ray  bursts (GRBs)  detected by the Burst and Transient
Source  Experiment  (BATSE) (Fishman et al.  1989) onboard the {\it Compton
Gamma-Ray  Observatory  (CGRO)} is a few times larger than the yield of all
other  experiments  that have detected GRBs.  Despite that the breakthrough
in 1997  concerning  the distance to GRBs was  associated  with the precise
localization of a {\it small} number of GRBs by {\it  Beppo-SAX}, the value
of the {\it large} BATSE sample cannot be  overestimated.  Problems where a
large statistics of GRBs as well as the wide brightness  range of the BATSE
sample are crucial are: \\
1. estimating the total rate of GRBs in the Universe, \\
2. searching for different subclasses of GRBs, \\
3. observing gravitational lensing of GRBs,  \\
4. searching for repetition of GRBs, \\
5. estimating the intrinsic luminosity function of GRBs. \\
The list can be made longer. Here, we mention only those problems which are 
directly associated with the goals of this work.

Nevertheless,  the BATSE sample (i.e., the sample of the  triggered  events
included in the BATSE  catalogs\footnote[5]{The  BATSE catalog is available
at http://www.batse.msfc.nasa.gov/batse/}) is smaller than what it could be
at the actual  sensitivity  of BATSE.  The reason  for this is a  difficult
variable  background.  The trigger threshold must be higher and the trigger
integration  time  must  be  shorter  than  for  the  case  of  a  constant
background.  Otherwise a high rate of  triggers  would have  caused  severe
problems  in the  data  acquisition.  The  BATSE  trigger  adjusted  to the
background  conditions missed many weak but still highly  significant GRBs.
Some bursts were also missed due to other  reasons  (data  readouts,  large
background, etc.).

Many of these non-triggered GRBs can be confidently identified in the BATSE
daily  records  which cover the whole  period of the {\it CGRO}  operation.
Off-line searches for non-triggered bursts with less rigid off-line trigger
criteria can substantially increase the sensitivity of BATSE and extend the
sample of GRBs both in number and in brightness range.

The  first  sample  of  non-triggered  GRBs was  published  by Rubin et al.
(1992).  A  systematic  search  for  non-triggered  GRBs in 6 years of {\it
CGRO}/BATSE    data   was    performed    by   Kommers    et   al.   (1997,
1998,\footnote[6]{The catalog is available at  http://space.mit.edu/BATSE/}
2000,  hereafter  K97, K98, K00).  Schmidt  (1999)  performed a scan of the
BATSE daily records for $\sim$6 years (TJD 8365--10528)  using a triggering
procedure  close to the BATSE trigger and found about 400 bursts  missed by
BATSE itself.

The main feature of our off-line scan of the BATSE data is the  measurement
of the  efficiency of the GRB detection as a function of  brightness  using
artificial  test bursts (e.g., Stern et al.  2000b).  This method plays the
role  of  being  a  calibration  of  the  experiment.  

In this paper, we concentrate  on a description  of the data  reduction and
the  tests  for  possible  non-GRB  contaminations.  We also  describe  the
publically  available  data base  containing  our  sample of GRBs.  Earlier
preliminary  versions of this data base and some first results were briefly
reported in Stern et al.  (1999a, 1999b, 2000a, 2000b).

 In \S\S 2--5, we describe the  procedures of the scan, the  identification
of GRBs,  and the GRB data  reduction.  In \S 6, we  describe  the  general
characteristics  of our GRB sample,  compare the sample  with the BATSE and
the Kommers et al.  catalogs, and describe the data archive  containing the
sample.  Section 7 is devoted to  various  statistical  tests for  possible
non-GRB   contaminations   using  hardness  ratios,  angular  and  latitude
distributions.  We obtain  upper  limits to the  contaminations  caused  by
terrestrial  phenomena, solar flares, and Galactic X-ray objects.  In \S 8,
we estimate the efficiency of the scan and present the final $\log N - \log
P$ distribution, which considerably differs from those of the BATSE and the
Kommers et al.  catalogs.

\section{The Data Scan} 

We used the  BATSE  daily  records  (DISCLA)  from the ftp  archive  at the
Goddard      Space      Flight       Center.\footnote[7]{Available      at:
http://cossc.gsfc.nasa.gov/pub/data/batse/daily/} The main array of the raw
data  consists  of the number  counts in the 8 BATSE  Large Area  Detectors
(LADs) in the 4 energy channels with 1.024 s time  resolution for the whole
period of observations (excluding some data gaps).

Examples of data  fragments are presented in  Figure~\ref{fig:1}.  The data
are  difficult to process  because of the large  diversity  of  interfering
phenomena:  the variable  background, the flaring ionosphere, solar flares,
variable  astrophysical  sources, and  occultations of sources by the Earth
causing  steps in the count rate curves.  In  addition,  the  situation  is
confused by numerous data gaps.

\medskip
\centerline{\epsfxsize=8.5cm {\epsfbox{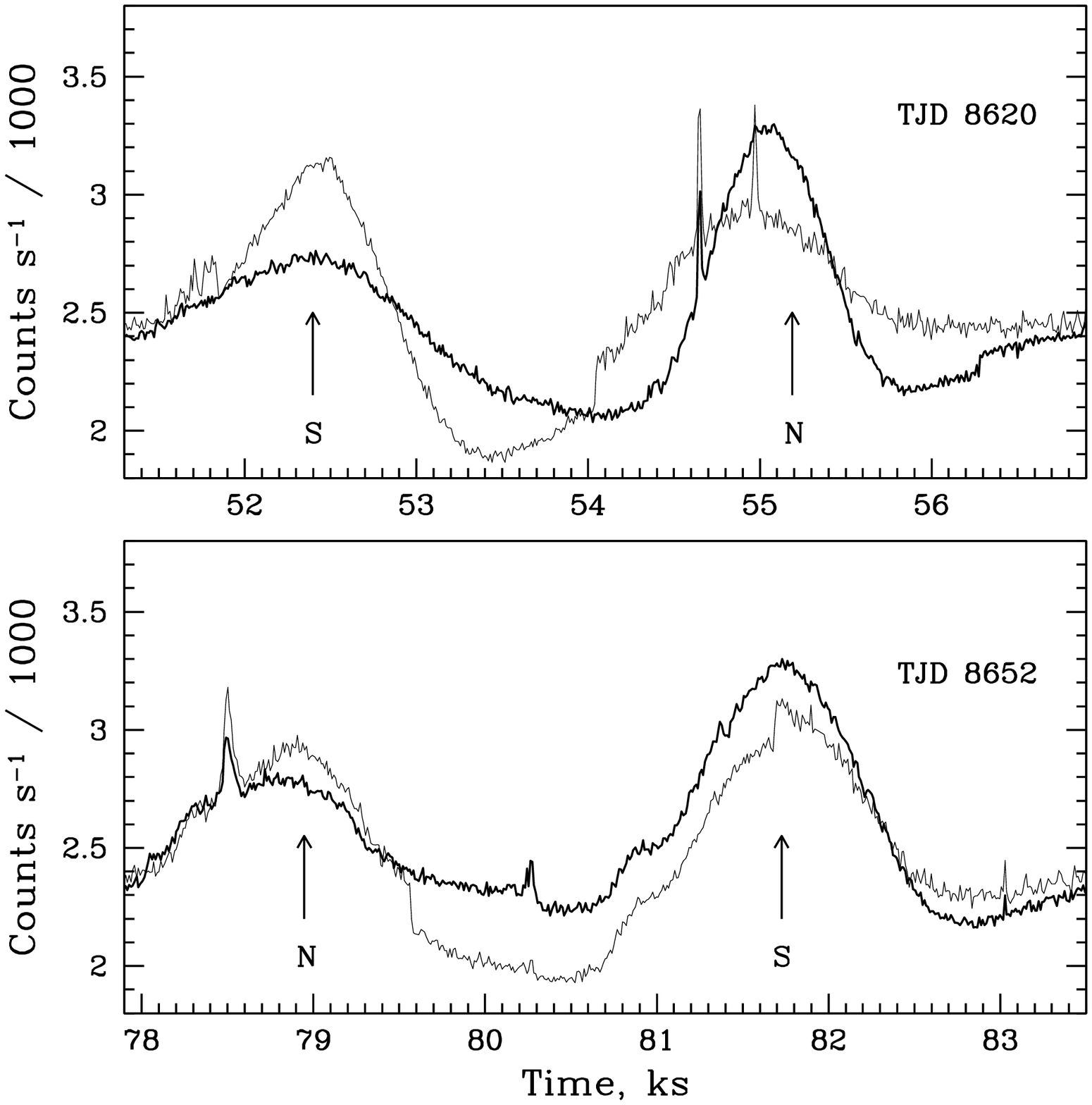}} }
\figcaption{
Two fragments of the BATSE daily records, each corresponding to one orbital
period of {\it CGRO}.  Counts rates  averaged over ten 1.024 bins are shown
for 2  detectors  in the  50--300  keV  range.  Features  of the count rate
curves are:  (Upper  panel) $<$51 800 s -- strong  noise from Cyg X-1 which
is in its bright  hard  state; 51 800 s --  descent of Cyg X-1; 52 400 s --
{\it CGRO} reaches its most Southern latitude; 54 050 s -- rise of Cyg X-1;
54 660 s -- particle  precipitation visible in all 8 detectors; 54 930 s --
a solar flare; 55 200 s -- {\it CGRO} reaches its most  Northern  latitude;
56 270 s -- rise of the Crab  nebula.  (Lower  panel)  78 500 s -  particle
precipitation;  79 550 s -- descent of Cyg X-1; 80 250 s -- a non-triggered
GRB (the brightest and the third brightest  detectors); 81 700 s -- rise of
Cyg X-1; 83 030 s -- a  non-triggered  GRB (the two  brightest  detectors).
This is a rare case of observing two  non-triggered  GRBs during one orbit.
Both  are  very  confident.  They  have  peak  fluxes  of  0.14  and  $0.21
\ph\secinv\cminvsq$, respectively.
\label{fig:1}
}
\medskip

The  complicated  background  makes  usual  statistical  estimates  of  the
detection   efficiency  useless.  The  estimate  derived  considering  only
Poisson  fluctuations  does not work when  non-Poisson  variability  of the
background  is strong.  Then there are a variety of reasons  for  missing a
detectable burst:  a data gap, interference with a variable source, or high
ionospheric  activity.  The  corresponding  probabilities  are in all cases
brightness  dependent.  In this case, a realistic estimate of the detection
efficiency should be based on some kind of simulation of the real detection
conditions  using real data.  Such a simulation was implemented in the form
of a test burst method (Stern et al.  2000b).

Artificial  test bursts were prepared  from a sample of 500 real  triggered
BATSE bursts of  durations  longer than 1 s and were added to the  original
``daily''  count rate records.  Each test burst was created by sampling one
of the 500 bursts with a random  number and  rescaling  its  amplitude to a
randomly sampled expected peak count rate with a proper Poisson noise.  The
locations of the test bursts were sampled  isotropically over the sky above
the Earth horizon, and their arrival times were  distributed  randomly with
an average  interval  $\sim25000$  s (in total,  there  were $11\ 100$ test
bursts).  The  distribution  of peak  count  rates  prescribed  to the test
bursts is shown in  Figure~\ref{fig:2}.  The shape of the  distribution  is
complicated  because  initially we did not have a clear  optimal  scheme of
brightness  sampling of test  bursts.  Initially,  we sampled the  expected
peak count rate  between  $0.08  \counts\secinv\cminvsq$  and $0.3  \counts
\secinv\cminvsq$  (uniformly  in  logarithmic  scale)  for 60\% of the test
bursts.  This  interval  was  believed to cover the main  variation  of the
efficiency  curve and no detections  below $0.08 \counts \secinv  \cminvsq$
were  expected.  The remaining  40\% of the test bursts were sampled in the
brightness  interval $0.3 - 3.0 \counts \secinv  \cminvsq$  (according to a
power-law $dN/dP \propto  P^{-1.6}$) in order to check whether there exists
any  efficiency  versus  brightness  dependence  for bright GRBs.  Later we
improved  the  efficiency  of the search and found that we detect some real
events below the above threshold.  Therefore, we reduced the lower limit to
$0.06 \counts \secinv \cminvsq$ and later to  $0.05\counts\secinv\cminvsq$.
As a result, we have an artificial  and  non-optimal  distribution  of test
bursts.  However, it is still suitable for estimating the efficiency.

\medskip
\centerline{\epsfxsize=8.5cm {\epsfbox{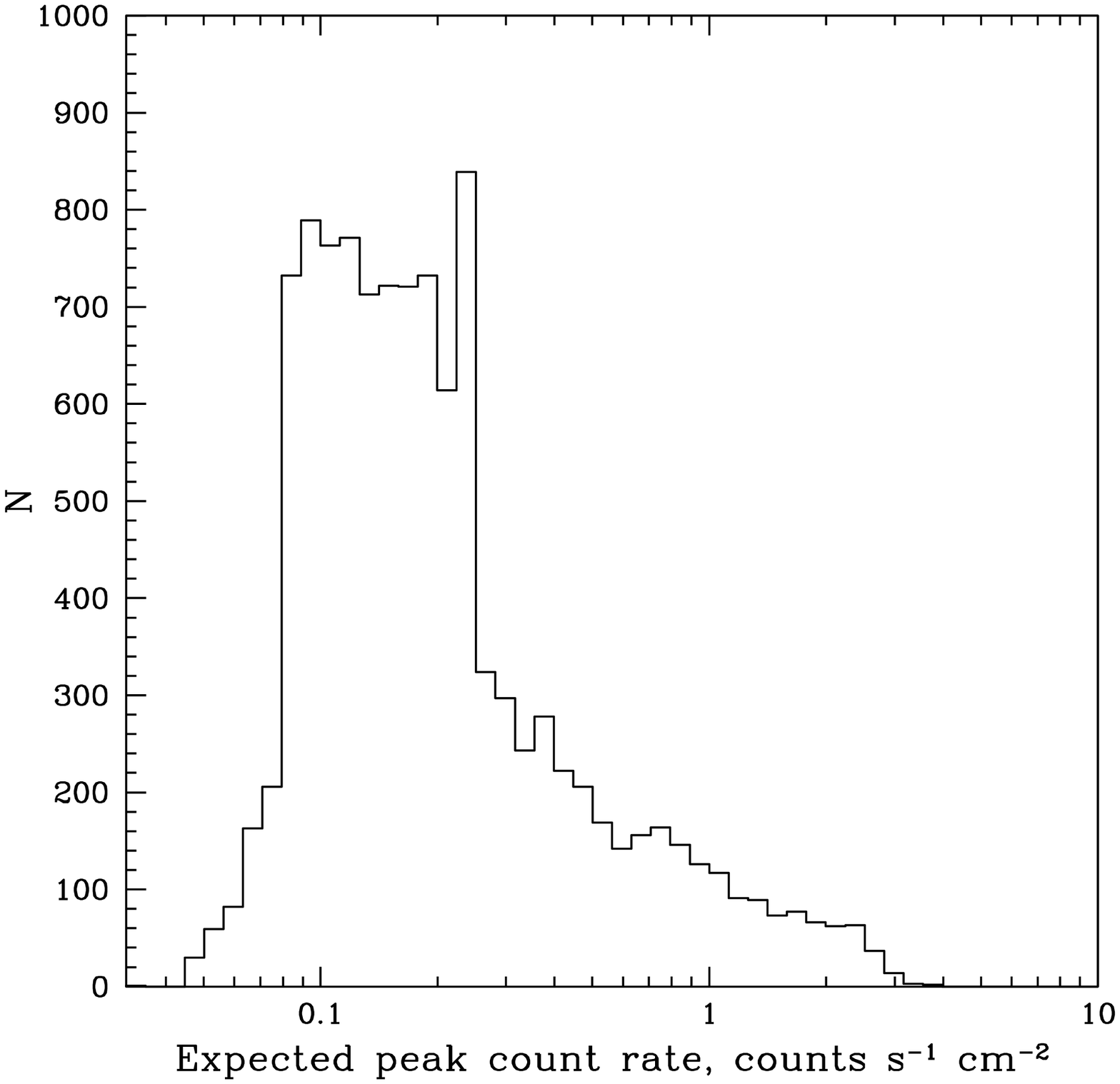}} }
\figcaption{
Peak  count  rate  distribution  of test  bursts  added  to the  data  (for
transformation of these units to $\ph \secinv \cminvsq$ or to total counts,
see \S~5.2).  Test bursts are included here  independently  of whether they
were  detected  or missed.  The BATSE  threshold  is around  $0.15  \counts
\secinv \cminvsq$.  The shape of the distribution is discussed in the text.
\label{fig:2}
}
\medskip

The  procedure of the scan is  described  in Stern et al.  (1999b,  2000b).
Briefly, it consists of adding test bursts to the data, the  triggering, an
express-analysis, the identification, and the classification.

The  off-line   trigger   consists  of  three   criteria  to  be  satisfied
simultaneously.  The first is a traditional one:  a significant rise in the
counts  over  the  estimated  background.  For the  brightest  detector,  a
$4\sigma$ excess was required, and for the second  brightest, a $2.5\sigma$
excess.  This is lower than the BATSE threshold and similar to that of K00.
The excess was checked for in time intervals (i.e., triggering time scales)
of 1 bin  (1.024  s), 2 bins, 4 bins,  and 8 bins.  The  longer  triggering
integration  time scales gives the main gain in the sensitivity as compared
to BATSE.  The second criterion is a sufficiently  high variability  around
the trigger time.  The  variability  is expressed as the residual  $\chi^2$
after  making a linear fit of the signal.  This  $\chi^2$  should  exceed a
threshold  value,  2.5 per  degree of  freedom.  This  value  was  adjusted
empirically  in a trial scan.  The third  criterion is a check  whether the
signal can be attributed to Cyg X-1.  The signal was fitted with the photon
flux  from  the Cyg X-1  direction  and  the  residual  was  checked  for a
sufficient  variability as in the previous case.  The two latter  criteria,
which  were  not  used  previously,  turned  out to be  very  efficient  in
eliminating  false triggers  reducing their number by more than an order of
magnitude.

Each trigger was followed by a preliminary  estimate of whether the trigger
is a GRB candidate (the person  performing  the scan was unaware of whether
the  candidate  event  was a real  or a test  burst).  Most  triggers  were
rejected  by a visual  analysis.  Usually  false  triggers  were  caused by
occultation steps, sharp variations of the ionospheric background, or solar
flares.  The  selection  criteria  used at this step are  similar  to those
described in \S 4, they are just not as  stringent,  and the count rate was
just assumed to be  proportional  to the  detector  projection  area in the
burst  direction  ($\propto  \cos  \theta$).  For all  selected  events, we
recorded  fragments of the original data and some preliminary  estimates of
their parameters.

At the next step, we identified  which recorded  events were test bursts or
BATSE triggers.  Finally, we processed all recorded  events and made a more
careful selection of GRBs using the procedures  described in \S 3 and \S 4.

The  rate  of  our  off-line  detections  of  GRBs  depends  on  time  (see
Fig.~\ref{fig:3}).  It reflects the quality of the data, the  intensity  of
noise-generating  sources (Cyg X-1, X-ray  novae), and the solar  activity.
One can see a gradual  increase in the number of  detected  GRBs with time.
It results from improvements in the software and from the experience gained
during  the scan.  Note that our rate of  observing  non-triggered  GRBs is
well  correlated  with that of K98.  The scan  itself  took for our group 2
years.  The speed of the scan  gradually  increased.  Now the primary  scan
would take about 200 full  working  days for one  person  and a  comparable
amount of work would be spent on the second stage data processing.

\section{The Fitting of the GRB Location and Photon Flux} 

 The aim of the fit is to find  the  best  burst  location,  spectrum,  and
intensity  describing  the  count  rate  data in 8  detectors  and 4 energy
channels.  This is done using the detector response matrix (hereafter DRM),
$D_{ljk}$,  where $l$ is the  index of the  photon  energy  bin, $k$ is the
channel number, and $j$ is the detector  number.  $D_{ljk}$  depends on the
burst  location, the satellite  orientation,  and the location of the Earth
with respect to the {\it CGRO} coordinate  system.  To calculate the DRM we
used a version of the code written by G.  Pendleton  which was used for the
localization  of  the  BATSE  triggered  bursts  (Pendleton  et  al.  1999,
hereafter  P99).  The  localization  procedure  itself  differs  from  that
implemented in the LOCBURST code by P99 in some details as described below.

\medskip
\centerline{\epsfxsize=8.5cm {\epsfbox{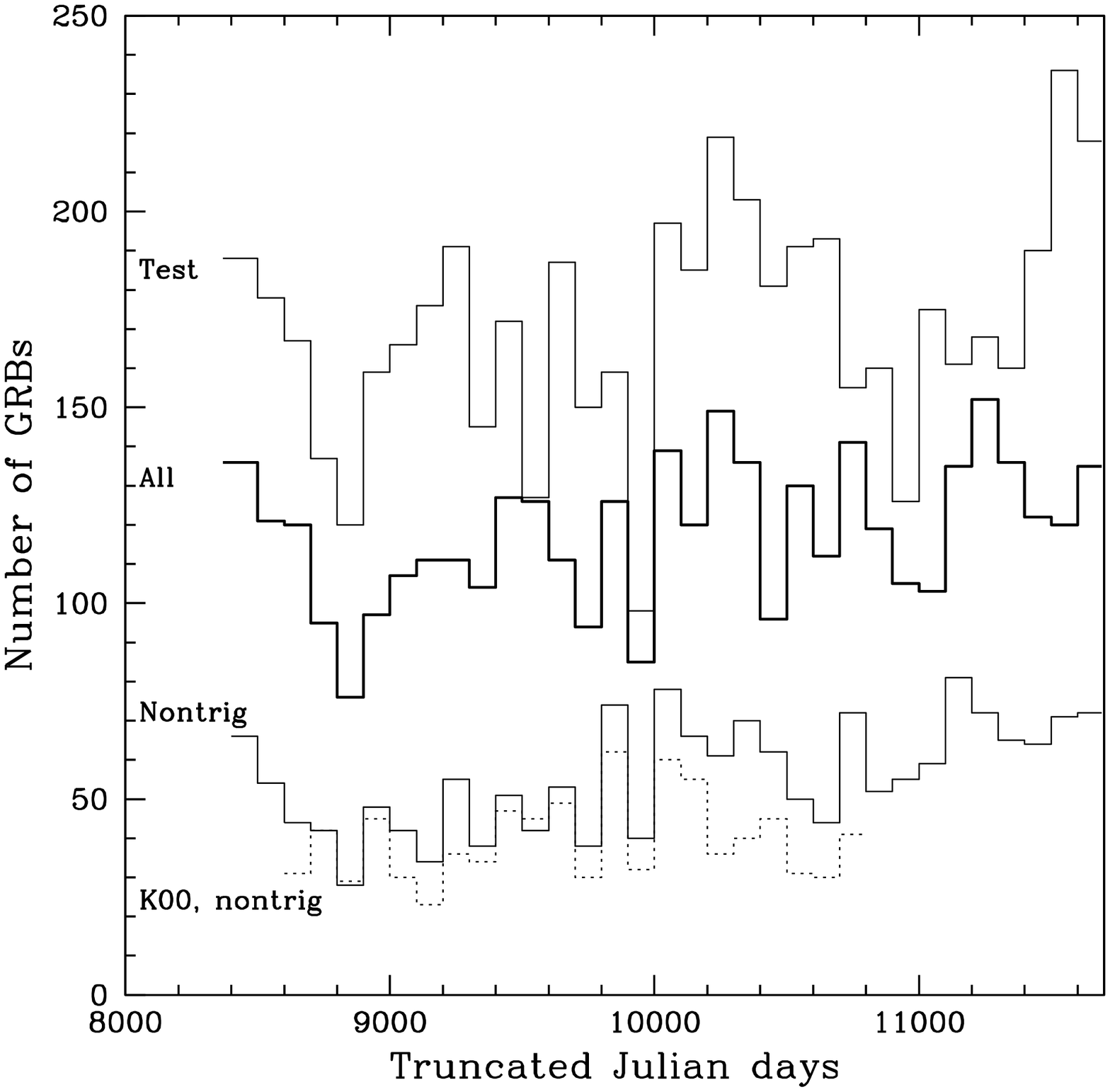}} }
\figcaption{
Time history of the rate of GRB detections during the scan for (from top to
bottom)  test   bursts,  all  real   bursts,   non-triggered   bursts,  and
non-triggered  bursts of K00.  The two deepest minima at TJD 8800--8900 and
9900--10000 are caused by bright X-ray novae.
\label{fig:3}
}
\medskip
 As a spectral  hypothesis, we used the Band  parametrization  (Band et al.
1993),  which  consists of 4  parameters:  the low energy  power law slope,
$\alpha$; the high energy power law slope,  $\beta$; the transition  (peak)
energy, $E_0$; and the  normalization  factor.  The problem is that we have
only  4  experimental   values:  the  counts  in  the  4  energy  channels.
Therefore  a   straightforward   fit  using  the  Band  function  would  be
degenerate.

We solved this problem using a sample of Band parametrizations (Band et al.
1993) of real  time-integrated  spectra of 54 GRBs.  At the first  step, we
determined  which of the 54 spectra (i.e., which  combination  of $\alpha$,
$\beta$  and  $E_0$)  that  provides  the best  fit of the  time-integrated
brightest  part of the burst  using a  pre-estimated  burst  location.  The
location fit used by P99 assumed that the GRB spectra are power laws (i.e.,
a 2-parameter  description),  but this is a worse  description  of real GRB
spectra as compared to the Band function.

For a given spectrum $F_l$ we have 
\begin{equation} \label{eq:cijk}
C_{ijk} = {\sum_l D_{ljk} F_l I_i /{F_{50-300}}} = d_{jk} I_i,
\end{equation}
where $C_{ijk}$ is the net GRB signal in counts in the $i$-th time bin, the
$j$-th  detector,  and  the  $k$-th  energy  channel,  $F_{50-300}$  is the
integral of the incident  photon number  spectrum in the 50--300 keV energy
range, and $d_{jk}$ is the  spectrum-integrated  detector  response matrix.
In this  normalization,  $I_i$ is the  photon  flux in units  of  $\ph
\secinv \cminvsq$.  The value to be minimized is
\begin{equation} \label{eq:chi2}
\chi^2 = \sum_{ijk}  \frac{(d_{jk} I_i + B_{ijk} - S_{ijk})^2} {S_{ijk}},
\end{equation}
where $B_{ijk}$ is the estimated  background for time bin $i$ and $S_{ijk}$
is the measured  count rate.  The  following  parameters  are free to vary.
For the whole event there are three parameters:  two parameters to describe
the burst  location  (the  factor  $d_{jk}$  depends on  location)  and one
parameter  for the choice of the spectrum  from the Band  sample.  For each
time bin there are two more  parameters:  the  photon  flux  $I_i$  and the
spectral  peak energy  $E_0$ (at this stage, we kept  $\alpha$  and $\beta$
fixed at their best-fit  values).  Having two parameters  ($I_i$ and $E_0$)
free at every time bin is  statistically  meaningful  and our chosen  model
describing  the GRB  spectral  evolution  as a  variation  of  $E_0$  seems
natural.

The background is estimated as a linear function of time, independently for
each detector and each energy channel using two fitting windows, one before
and one after the main peak of the event.  These  windows were set manually
using a visual interpretation of the count rate time profile.  In some rare
difficult cases, only one background fitting window was used.

The value of $\chi^2$ was calculated  for a few thousand  locations  over a
wide area of the sky.  Using the  $\chi^2$  map of the sky, we defined  the
best fit location, the $1 \sigma$  error  boundary,  and the quality of the
event as a GRB  candidate  (see \S 4).  The  most  serious  problem  in the
location fit is the poor accuracy of the DRM for the detectors which do not
see the burst directly but still detect the burst due to photon  scattering
by the Earth's  atmosphere or due to penetration  through the satellite.  A
precise  calculation  of the DRM is an extremely  difficult  problem as one
should take into  account the 3D mass  distribution  of the {\it CGRO}.  In
the LOCBURST  algorithm, the final  localization  fit is performed only for
the 4 or 6 brightest  detectors.  This reduces the systematic error for the
best fit location.  

However, this option causes a problem when we make the  localization  for a
weak burst where a wide  $\chi^2$ map of the sky is needed (see \S 4).  For
different locations, different detectors become the brightest and it is not
easy to find the 4 or 6 brightest  detectors  from the data.  As far as our
main interest  concerns  weak events which  requires a large area  $\chi^2$
map, we preferred to use the fit for all 8 detectors.  Another advantage of
the 8  detector  fit is a  slightly  better  signal to noise  ratio for the
fitting net signal given by equation (\ref{eq:cijk}).

The systematic errors of our procedure is probably slightly worse than that
of LOCBURST.  However, the difference is considerable  only for very bright
bursts.  A more exact  localization  could be achieved,  in  principle,  by
iterations  using a  refined  DRM and  reducing  the  number  of  detectors
participating in the fit.  Since the exact localization of strong bursts is
beyond the scope of this work at its present stage, this was not done.  For
deviations between our and the BATSE locations, see \S 6.

The best fit value of the photon flux, $I_{i}$, for a given burst  location
and  photon  spectrum  can be  obtain  by  minimizing  the  expression  for
$\chi^2_i$  similar to equation  (\ref{eq:chi2})  with  respect to $I_{i}$,
i.e.  by solving $\partial \chi^2_i/\partial I_i=0$:
\begin{equation} \label{eq:Ii}
I_{i} = \frac{\sum_{jk} (S_{ijk}-B_{ijk}) d_{jk} / S_{ijk}}
{\sum_{jk} d_{jk}^2 / S_{ijk}} .
\end{equation}
The fitting  count rate (that we prefer to use instead of the photon  flux,
$I_{i}$,  for  estimating   the  peak   brightness)   is  then  defined  by
equation~(\ref{eq:cijk}), or, in units of $\counts \secinv \cminvsq$ as
\begin{equation} \label{eq:cik}
c_{ik}=  (d_{mk}/ s_m) I_i,
\end{equation}
where $m$ is the index of the  detector  having the  largest  projected
area $s_m$ in the burst direction.

The  quantities  used in this  paper to  represent  the time  profiles  
and the brightness of GRBs are: \\
1. The count rate in the 50--300 keV energy band and the $i$-th time bin, 
$c_i = c_{i2}+ c_{i3}\ \counts \secinv \cminvsq$, with 
$c_{i2}$ and  $c_{i3}$ given by equation (\ref{eq:cik}). \\
2. The peak count rate, $c\ \counts \secinv \cminvsq$, which is a function of
$c_i$,  see \S 5.2. This quantity is also denoted as $P$ (mostly on figures) 
following the tradition. \\
3. The net (fitting) count rate,  $C_{ik} = \sum_j  C_{ijk}\  \counts \secinv$, 
where $j$ is the detector number, see equation~(\ref{eq:cijk}), or the net 
count rate in the 50--300 keV energy band:  $C_i= C_{i2} + C_{i3}$. \\
4. The peak photon flux, $I\ \ph \secinv \cminvsq$, which is a function of 
$I_i$, see \S 5.2. 

\section{The Identification of GRBs} 

The main quantitative  criterion for selecting real events  associated with
astrophysical sources from the variations of the terrestrial  background or
other  fluctuations was based on the residual $\chi^2$ map of the sky.  The
same  criterion was used as the  significance  threshold.  Calculating  the
residual  $\chi^2$ as a function of the burst  location  (eq.  2) we define
the  ``1$\sigma$  area'',  i.e.,  a  region  where  $\chi^2  -  \chi^2_0  <
\sqrt{2N}$  where  $N$ is  the  number  of  the  degrees  of  freedom,  and
$\chi^2_0$ is the minimum value of $\chi^2$.  The location error  according
to our preference is the maximum  distance,  $\delta_1$,  from the best fit
location to the boundary of the 1$\sigma$  area.  Similarly,  we define the
4$\sigma$ region and the maximum  distance,  $\delta_4$,  from the best fit
location.  Our criterion which any event without any exception  should pass
to be considered as a GRB  candidate is  $\delta_4<  90^\grad$.  Note, that
$4\sigma$  is the excess of the {\it  residual}  $\chi^2$  for the fit at a
$90^\grad$  displacement.  The {\it statistical  significance}  of an event
(defined by the $\chi^2$  with respect to the  background)  is always above
$7\sigma$  when the  $\delta_4$  criterion  is  satisfied.  The  $\delta_4$
criterion  efficiently  rejects the fluctuations of the diffuse ionospheric
background.  Examples  of  events  which  have  $\delta_4$   close  to  its
threshold value are shown in Figures~\ref{fig:4} and~\ref{fig:5}.

Further  criteria were more  qualitative and subjective.  First, we checked
whether one of the variable  sources (the Sun, Cyg X-1, or one of the known
X-ray  transients  currently  active) was inside or close to the  $1\sigma$
area.  Then we applied  additional  requirements.  First, a large  hardness
ratio (number of counts in the 50--100 keV energy range  exceeding  that in
the 30--50 keV range),  especially  for the events close to the position of
the Sun.  Second,  for  events  close to the  positions  of the hard  X-ray
sources such as Cyg X-1 or X-ray transients, a large domination  (estimated
visually) of the event over the typical current  fluctuations of the source
was required (a large  domination over both Poisson and  non-Poisson  noise
was a general requirement).  Short soft events were identified as outbursts
of X-ray  pulsars  and their  locations  were  always  consistent  with the
location of one of the known X-ray pulsars.

 For each event we checked visually the signal in different  detectors with
the  requirement  that the  signal is  visible  in more than one  detector.
However, no requirements  for the  significance of the signal in the second
brightest  detector  were  applied -- the signal  should  just be  visually
recognized.

 Special  attention was paid to the ``context'' of the event:  a GRB should
be well  isolated  in order to be  accepted.  If a  persistent  noise  from
approximately  the same  direction  and of  comparable  intensity or a high
ionospheric  activity  were seen, the event was  discarded.  The  isolation
criterion is a  subjective  one to some extent as it is very  difficult  to
formalize.

 There were no ``good'' events with a localization  significantly below the
Earth's  horizon (all tests bursts were sampled  above the  horizon).  Such
events  were  only   marginally   significant   or  did  not   satisfy  the
$\delta_4$-criterion.  Therefore  there was no reason  to use the  location
relative to the Earth's horizon as an independent  criterion.  Nevertheless
events  located near the horizon  were  treated  more  suspiciously  if the
ionospheric activity was high.

\medskip 
\centerline{\epsfxsize=8.5cm {\epsfbox{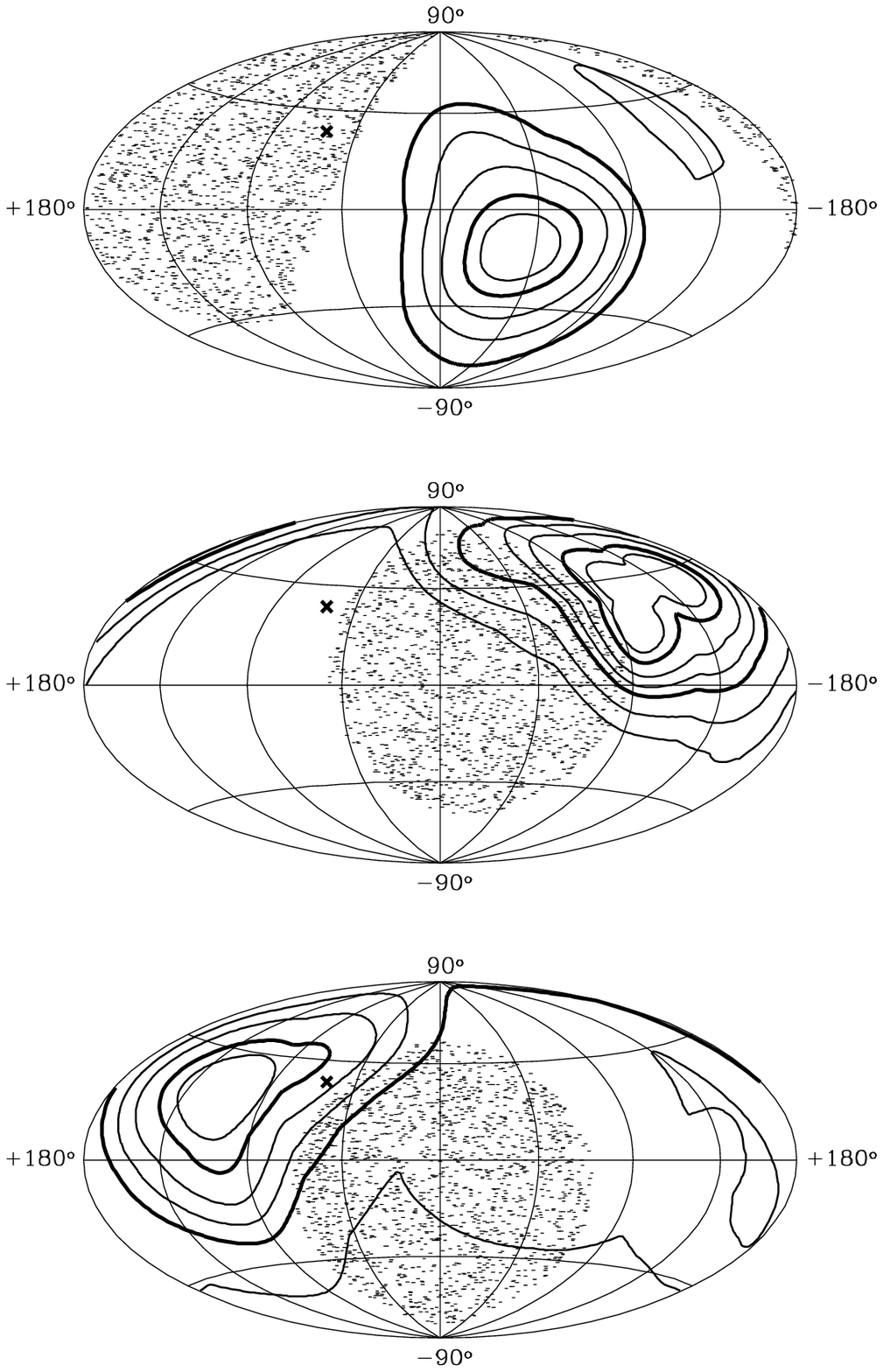}} }
\figcaption{ 
Residual  $\chi^2$ maps of the sky for three weak GRBs (09394b, 09575a, and
09590c, where the name consists of TJD and an  identifying  letter),  which
passed the selection  criteria being near the  $\delta_4$  threshold of the
significance (see text).  These events were chosen among the least reliable
cases.  The time profiles for these events are shown in Figure~\ref{fig:5}.
Isocontours  show  constant  levels of the  residual  $\chi^2$  measured as
$(\chi^2 -  \chi_0^2)/\sigma$  where  $\chi_0^2$ is the minimum,  $\sigma =
\sqrt{2N}$,  where $N$ is the number of the degrees of freedom.  The levels
at $0.5\sigma$,  $1\sigma$ (bold contour), $2\sigma$,  $3\sigma$, $4\sigma$
(bold  contour),  $6\sigma$,  and  $8\sigma$  are shown.  The  dashed  area
represents the Earth.  The cross shows the location of Cyg~X-1.
\label{fig:4}
}

\section{Estimating the Peak Brightness of GRBs} 

\subsection{The Choice of the Measure for the GRB Brightness} 

 There are two natural  choices for the  measure of the  strength  of GRBs.
One is a time  integral of the GRB signal:  the photon  fluence, the energy
fluence,  or  the  total  counts.  Another   option  is  to  use  the  peak
characteristics:  the peak photon  flux, the peak energy  flux, or the peak
count rate.  If we try to estimate the GRB distance distribution, a measure
which is closer to a standard  candle  (i.e., that has a smaller  intrinsic
dispersion)  would be  preferable.  We do not  know if this is the  case --
probably both the fluence and the peak flux have very large dispersions.

 The choice  is,  however,  natural  as the peak  characteristics  are much
easier to measure.  Besides, the detection  efficiency  depends on the peak
count  rate  rather  than on the  time-integrated  signal.  Among  the peak
characteristics  we choose  the peak  count  rate as a  measure  of the GRB
brightness  for the  following  reasons.  First, the peak count rate unlike
the peak  photon or energy  flux  does not  depend  on the  assumed  photon
spectra  which  could  be  rather   uncertain.  It  is  defined  by  direct
measurements.  Second, the detection efficiency is a direct function of the
peak count rate, but not of the peak photon  flux.  The  definition  of the
peak count rate when we have 8 detectors with different  orientations.  The
best choice at this step is the {\it fitting count rate}, {\bf $C_{ik}$} or
its reduced value {\bf $c_{ik}$} (see \S 3), i.e., the best fit  hypothesis
of the true signal.

\medskip
\centerline{\epsfxsize=8.5cm {\epsfbox{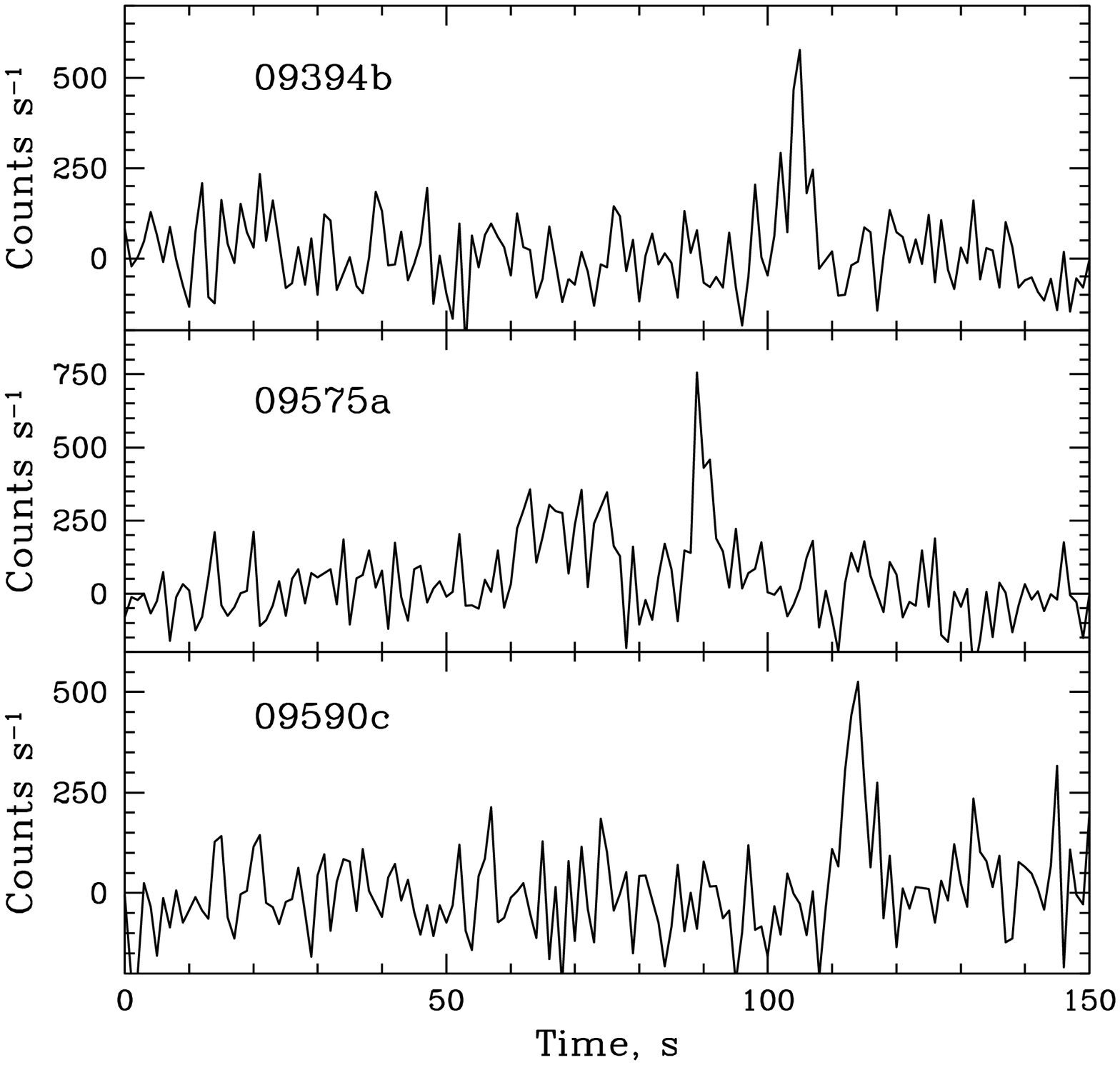}} }
\figcaption{
Time profiles for the 3 events that are shown in Figure~\ref{fig:4}.
Fitting count rates, $C_{i2}+C_{i3}$ (for channels 2+3, see section 3),
with subtracted background are given.
\label{fig:5}
}

\subsection{The Estimate of the Peak Count Rate} 

The  weakest  GRBs we deal with have a peak count rate of the same order of
magnitude  as  the  Poisson   fluctuations  at  1.024  s  time  resolution.
Therefore, if we define the peak count rate as the content of the brightest
time bin, we probably just choose the highest Poisson  fluctuation and thus
overestimate the brightness,  perhaps by a factor 2 or more for the weakest
events.  If we smooth the signal,  e.g.  by  averaging  over  several  time
bins, we reduce the Poisson  noise but then we may lose true short peaks of
the GRB signal.  A compromise  between  smoothing and  preserving  the time
resolution  is the  two-iteration  scheme  described in Stern,  Poutanen \&
Svensson (1999).

In the first  iteration, we find the shortest time scale, $\Delta T_j = 2^j
\times  1.024$~s,  $j=0, 1, 2, 3$, on which the  signal  has a  significant
variation  between  neighboring time bins (using a $4.3\sigma$  threshold).
Then we search for the  $\Delta  T_j$-interval  where the count rate is the
largest.  (If $\Delta T = 1.024$ s, the search is  completed.)  If $j > 0$,
we make a second iteration searching for a significant excess in a fraction
$\delta  T_l$  of  the  brightest  $\Delta  T_j$-interval  ($l <  j$).  The
significance  threshold,  $h_l$,  depends  on the time  scale  $2^l  \times
1024$~s  at this  step.  The  values of $h_l$  were  optimized  empirically
comparing  prescribed  (expected) and measured  count rates for test bursts
and were set to $h_0 = 3.5\sigma$, $h_1 = 2.4\sigma$, $h_3=1.4\sigma$.

 The  result of this  method  is shown in  Figure~\ref{fig:6}.  One can see
that a reasonable  linearity  between the expected and the  measured  count
rates has been  achieved.  Nevertheless,  after  getting rid of the Poisson
bias we still have a smaller bias of a similar origin.  Events  enhanced by
positive  Poisson  fluctuations  have a higher  probability to be detected.
Therefore we observe some systematic excess of the measured  amplitude over
the expected one for the weakest  bursts where the detection  efficiency is
low.  This bias is accounted for by the efficiency matrix (see \S 8.2).

\medskip
\centerline{\epsfxsize=8.5cm {\epsfbox{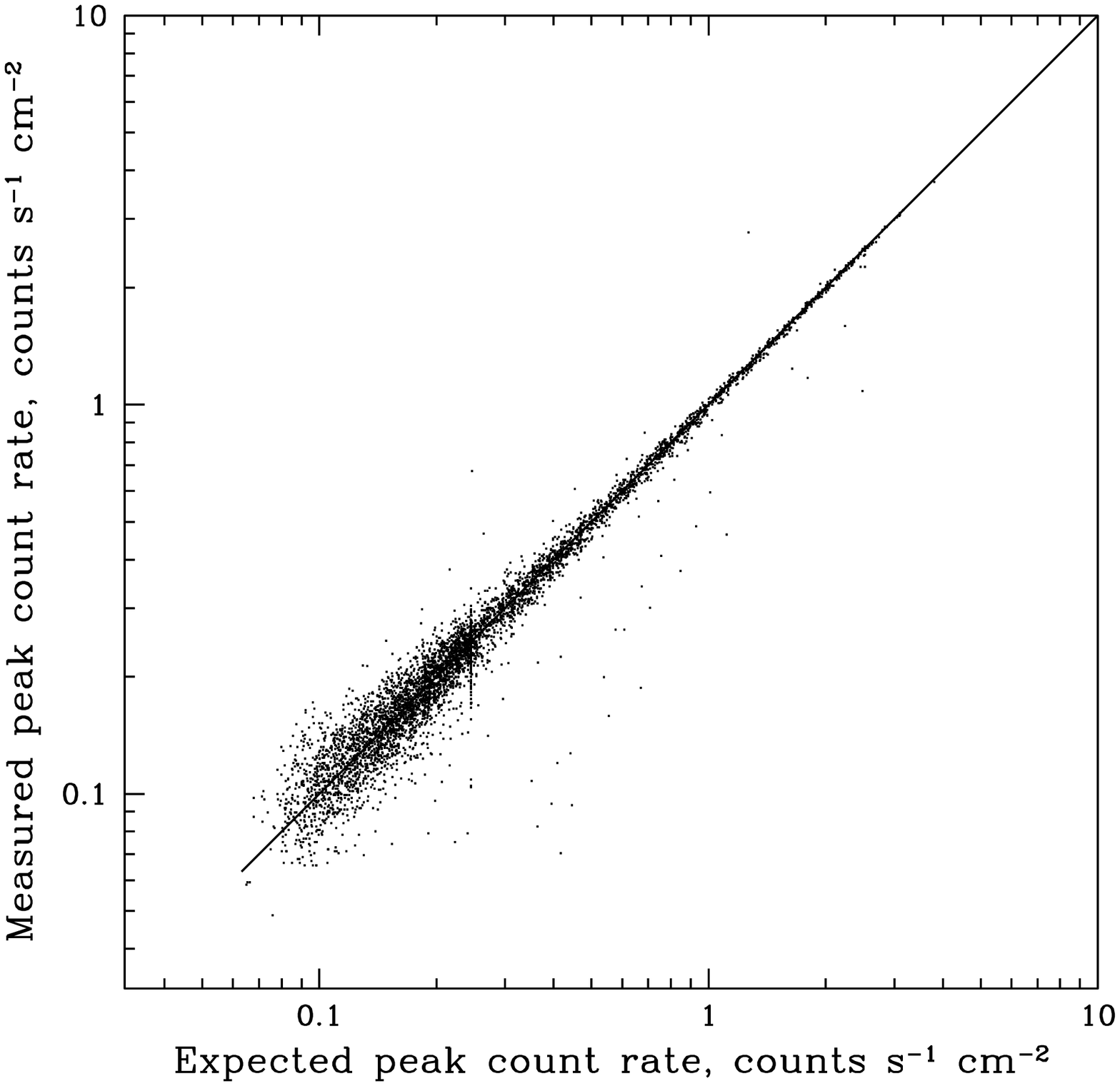}} }
\figcaption{ 
Measured vs expected  peak count rates for detected  test  bursts.  Several
points  having a measured  count rate much lower than the expected  one are
affected  by data  gaps  causing  the peak of the  event  to be  lost.  Two
points,  where the measured  count rate is much higher, are probably due to
confusion of the test burst with an overlapping brighter real event.
\label{fig:6}
}
\medskip

It is useful to give the approximate  ratio between the peak count rate and
the peak  photon  flux.  This ratio  versus the peak count rate is shown in
Figure~\ref{fig:7}.  There is no evident correlation  between the ratio and
the  brightness.  The  relation  between  the peak count  rate and the peak
photon flux can be  expressed  as $c= 0.7504\ I \pm  0.063$,  where the rms
variance  is given as the error.  Note that in rare cases the  50--300  keV
count rate  exceeds the 50--300 keV photon  flux.  The main reason for this
is the atmospheric  scattering of photons.  The photon flux by definition is
the flux of direct  photons,  while  the  photons  scattered  in the  Earth
atmosphere   contribute  to  the  count  rate.  For  some  geometries,  the
scattered component is considerable.  Another effect is that photons harder
than 300 keV can give a signal in the 50--300 keV range as they interact in
the detector via Compton  scattering.  Both effects enhance the 50--300 keV
count rate with respect to the photon flux in the same energy range and are
more pronounced for harder GRBs.

\section{The Sample and the Data Archive} 

We  performed  the scan of the DISCLA  data for the full 9.1 years of BATSE
observations up to the {\it CGRO} deorbiting (TJD 8369--11690).  The number
of  events  in our  sample  classified  as GRBs is  3906,  1838 of them are
non-triggered,  and 2068 we identified with BATSE triggers.  The peak count
rate  distributions  of these  GRBs are shown in  Figure~\ref{fig:8}.  Note
that some  non-triggered  events are very strong (the  strongest  one has a
peak flux of $24  \ph  \secinv  \cminvsq$).  The reason  for BATSE not
detecting some strong bursts is the dead time when the trigger was disabled
during data readouts or when {\it CGRO} passed through regions of very high
ionospheric activity.
On the other hand, a few brightest BATSE GRBs are missing from our sample
because their amplitude overflowed the two byte integer in the BATSE archive 
records and
we were not able to process them.

\medskip
\centerline{\epsfxsize=8.5cm{\epsfbox{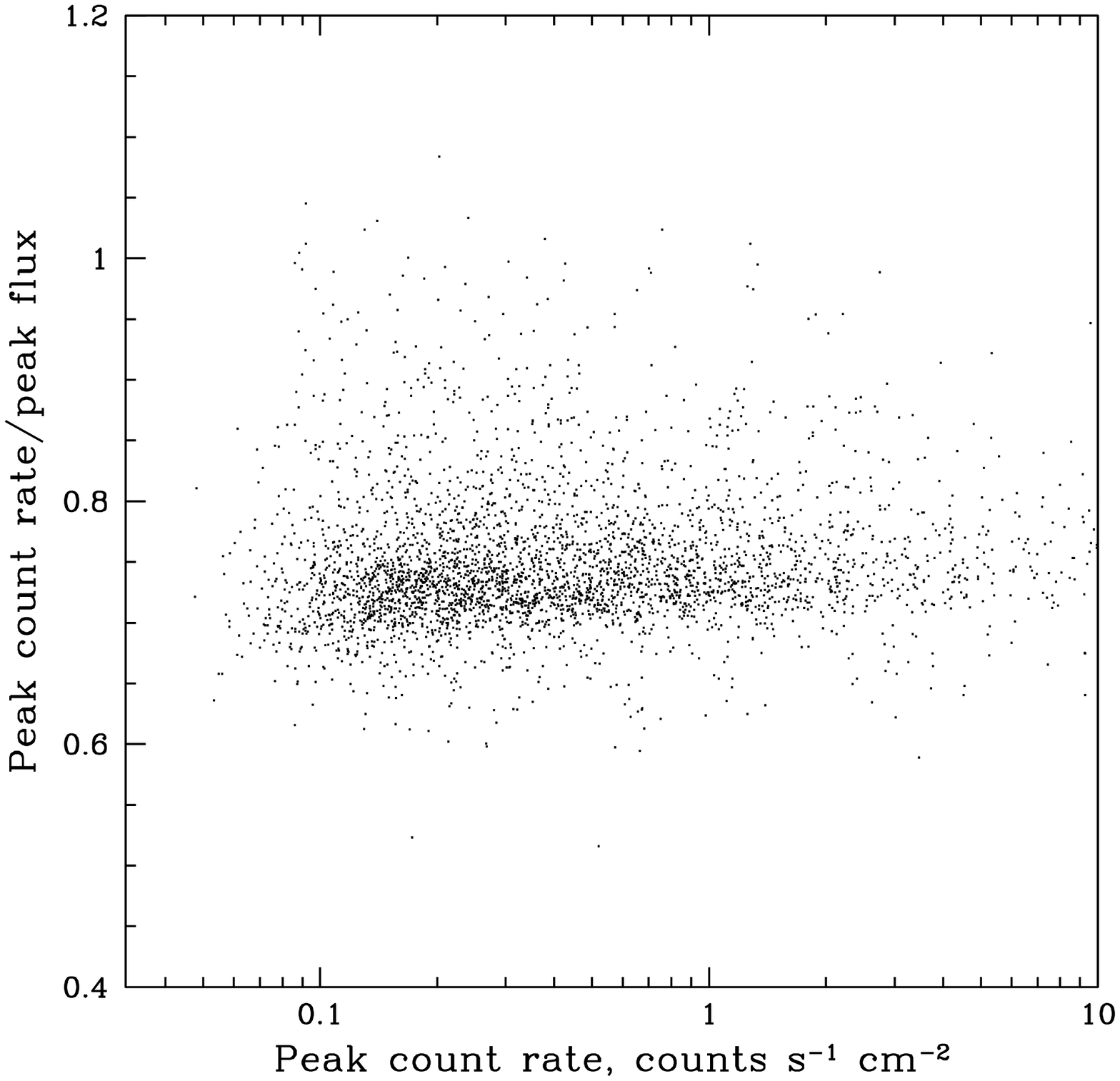}} }
\figcaption{ 
Ratio of the peak count rate in channels 2+3 ($\counts\secinv\cminvsq$) 
to the peak flux in the 50--300 keV band ($\ph\secinv\cminvsq$)  versus
the peak count rate.  A
few events with $c/I > 1$ are due to strong  atmospheric  scattering and to
the  hard  part of the  spectra  contributing  to  channels 2 and 3 
due to Compton scattering in the detectors.
\label{fig:7}
}
\medskip

Durations of GRBs were estimated using $T_{90}$ (Kouveliotou et al.  1993),
i.e., as the time  interval  between  the  emission  of 5\% and 95\% of the
total burst photon fluence.  The  measurement of $T_{90}$ has  considerable
uncertainties  when applied to weak bursts.  The result depends on the time
intervals  (windows)  where the signal and the background  are measured and
therefore   depends  on  the  subjective   impressions  of  the  researcher
estimating  what  is the  signal  and  what  are  the  fluctuations  of the
background.  The  duration   distribution   for  our  sample  is  shown  in
Figure~\ref{fig:9}.  One can see that  non-triggered  bursts  are  slightly
shorter on  average.  This can be  explained  by the  brightness  dependent
bias:  losses of some episodes of weak bursts.

\medskip
\centerline{\epsfxsize=8.5cm {\epsfbox{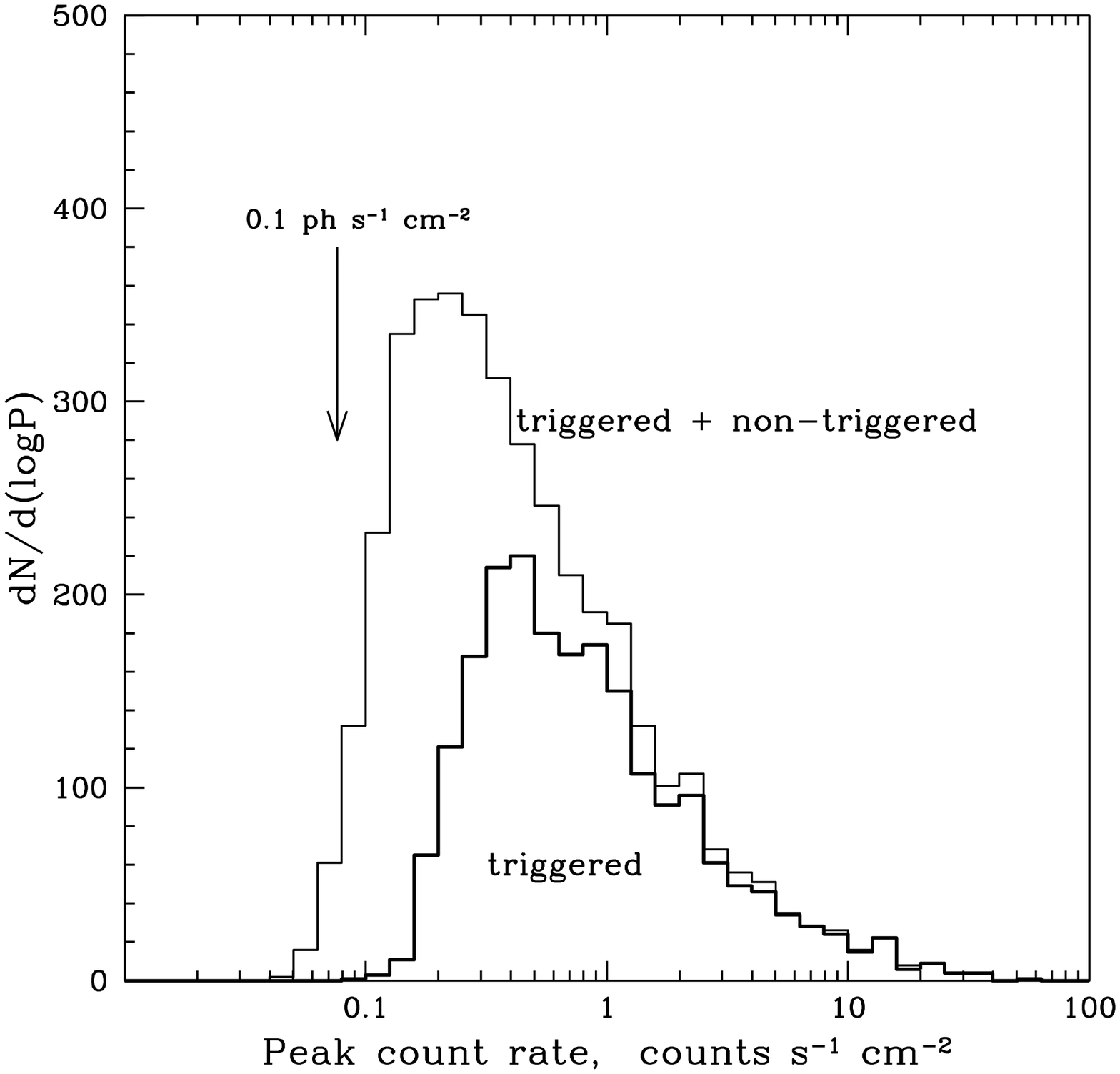}} }
\figcaption{ 
Peak count rate distribution,  $dN/d(\log P)$, of the 2068  BATSE-triggered
GRBs detected in our scan ({\it thick line histogram}) and of all 3906 GRBs
detected in our scan ({\it thin line histogram}).  The  distributions  have
not been corrected for the efficiency.
\label{fig:8}
}
\medskip

\medskip
\centerline{\epsfxsize=8.5cm {\epsfbox{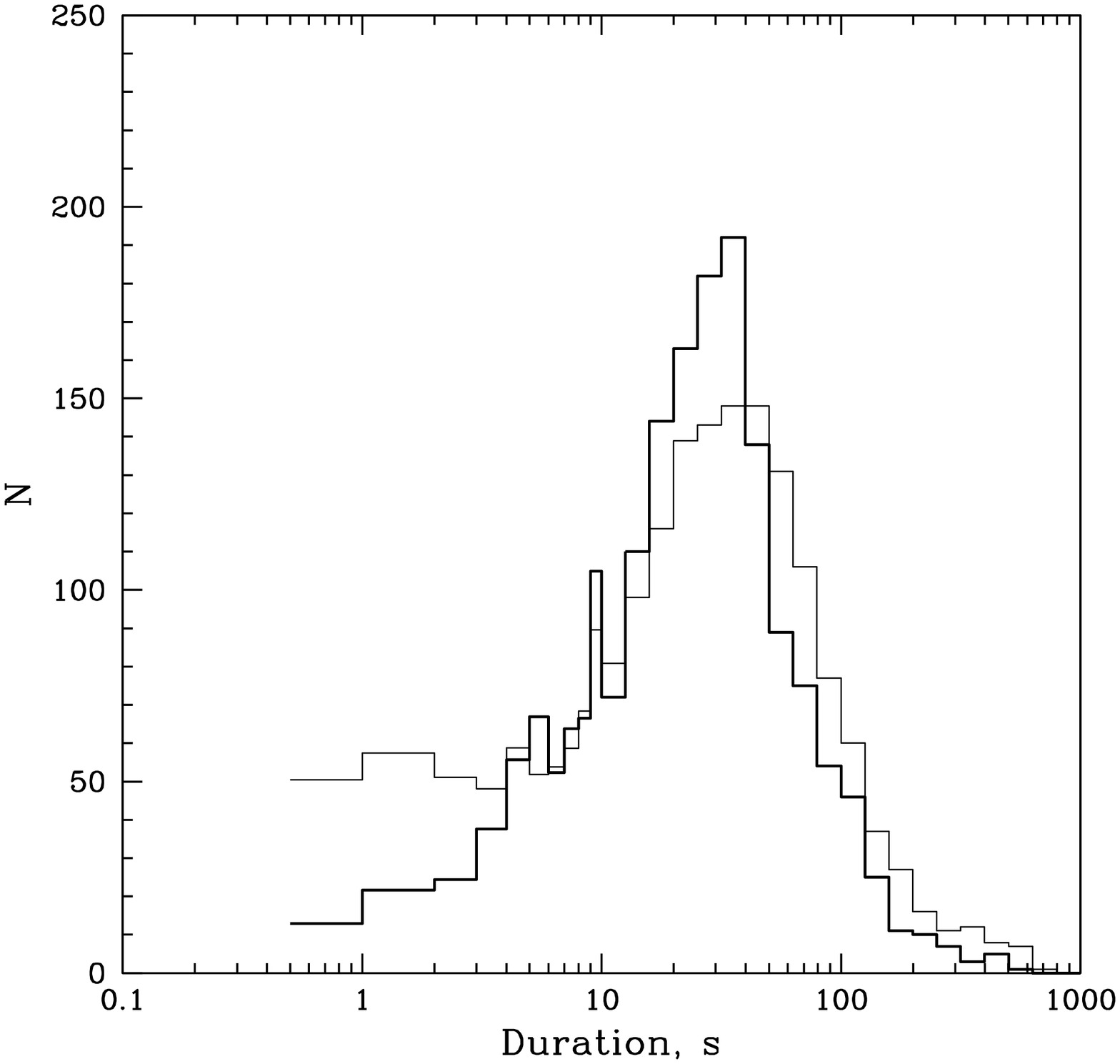}} }
\figcaption{
Duration  distribution  ($T_{90}$)  for  1965  triggered  ({\it  thin  line
histogram})  and 1713  nontriggered  ({\it  thick line  histogram})  bursts
detected up to TJD 11499.  The distribution at $T_{90} <4$~s is biased (see
text and Fig.~\ref{fig:10}).
\label{fig:9}
}
\medskip
\centerline{\epsfxsize=8.5cm{\epsfbox{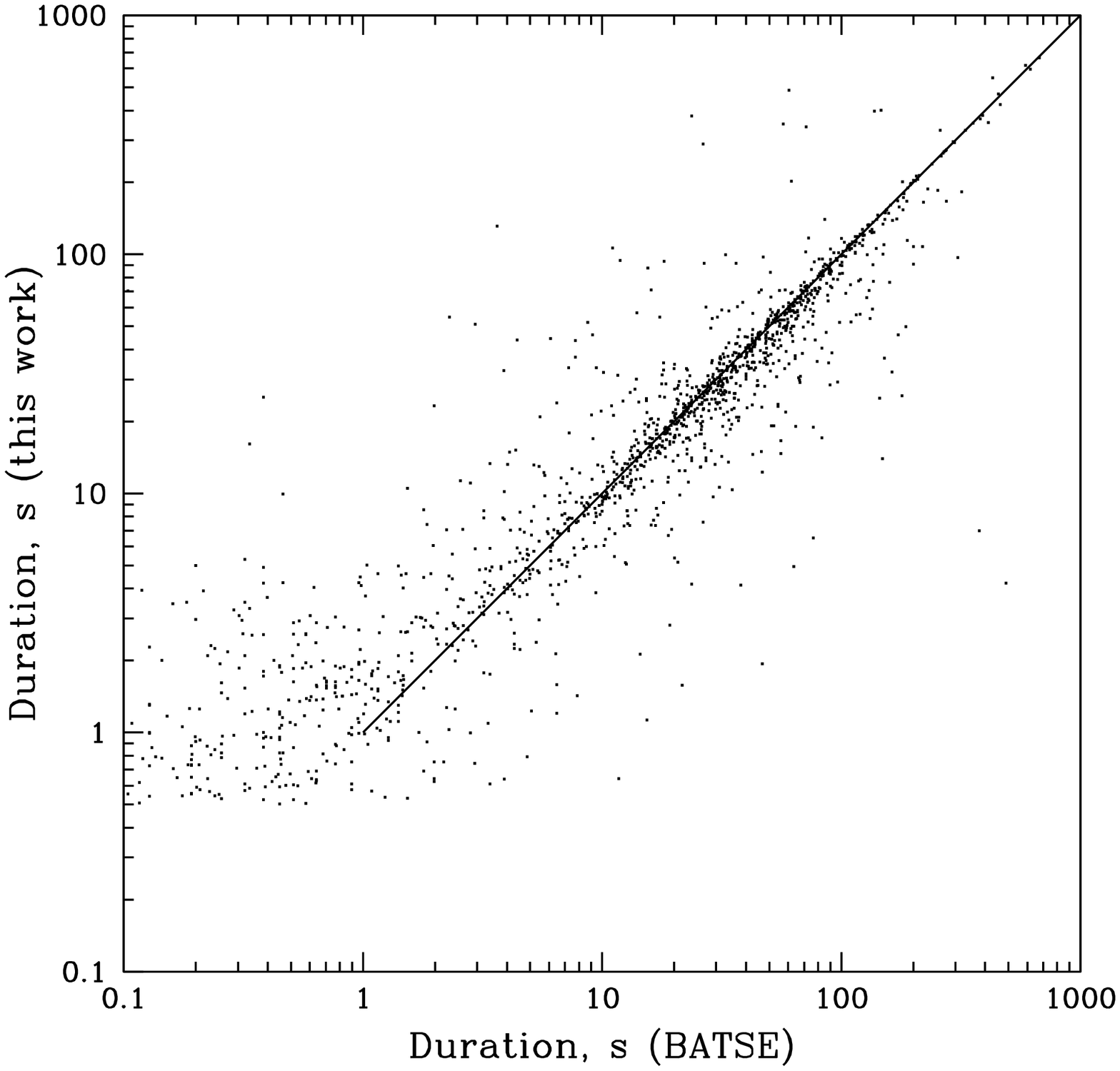}} }
\figcaption{
Comparison  between the BATSE and our  estimates  of $T_{90}$  for the 1965
bursts  detected  both by BATSE and by us up to TJD 11499.  For clarity, to
avoid the 1.024 s  discreteness  of the data, our  estimates  are dispersed
within 1.024 s bins (otherwise points would merge into horizontal lines).
\label{fig:10}
}
\medskip

The relation  between our and the BATSE  estimates  of $T_{90}$ is shown in
Figure~\ref{fig:10}.  Large   deviations   are   associated   with   events
consisting  of  widely  separated  episodes  with  significantly  different
brightnesses.  One of the two teams can easily  lose the  weakest  episode.
Another  problem  is  poor  precision  in the  estimation  of  the  fluence
distribution  as it depends on the details of the  background  fitting.  If
there is a precursor or an aftercursor  with the fluence around 0.05 of the
total one, then one team can include this episode into  $T_{90}$, the other
team can  exclude  it.  Note,  that  deviations  between  our and the BATSE
estimates are of both signs and approximately symmetric for long GRBs.  Our
estimates of $T_{90}$ for short events are strongly biased due to the 1.024
s resolution  and the Poisson  fluctuations.  There is a large  probability
that an event with $T_{90} < 1$~s will be distributed in two time bins (and
therefore  will be  estimated  as 2 s long)  or will  merge  with a  nearby
Poisson fluctuation.

The  distribution  of the best fit locations of the GRBs in our sample over
the sky is shown in  Figure~\ref{fig:11}.  No significant  deviations  from
the   BATSE   exposure   function   (see   Paciesas   et   al.  1999;   and
Figure~\ref{fig:17}) have been found.

\medskip
\centerline{\epsfxsize=8.5cm {\epsfbox{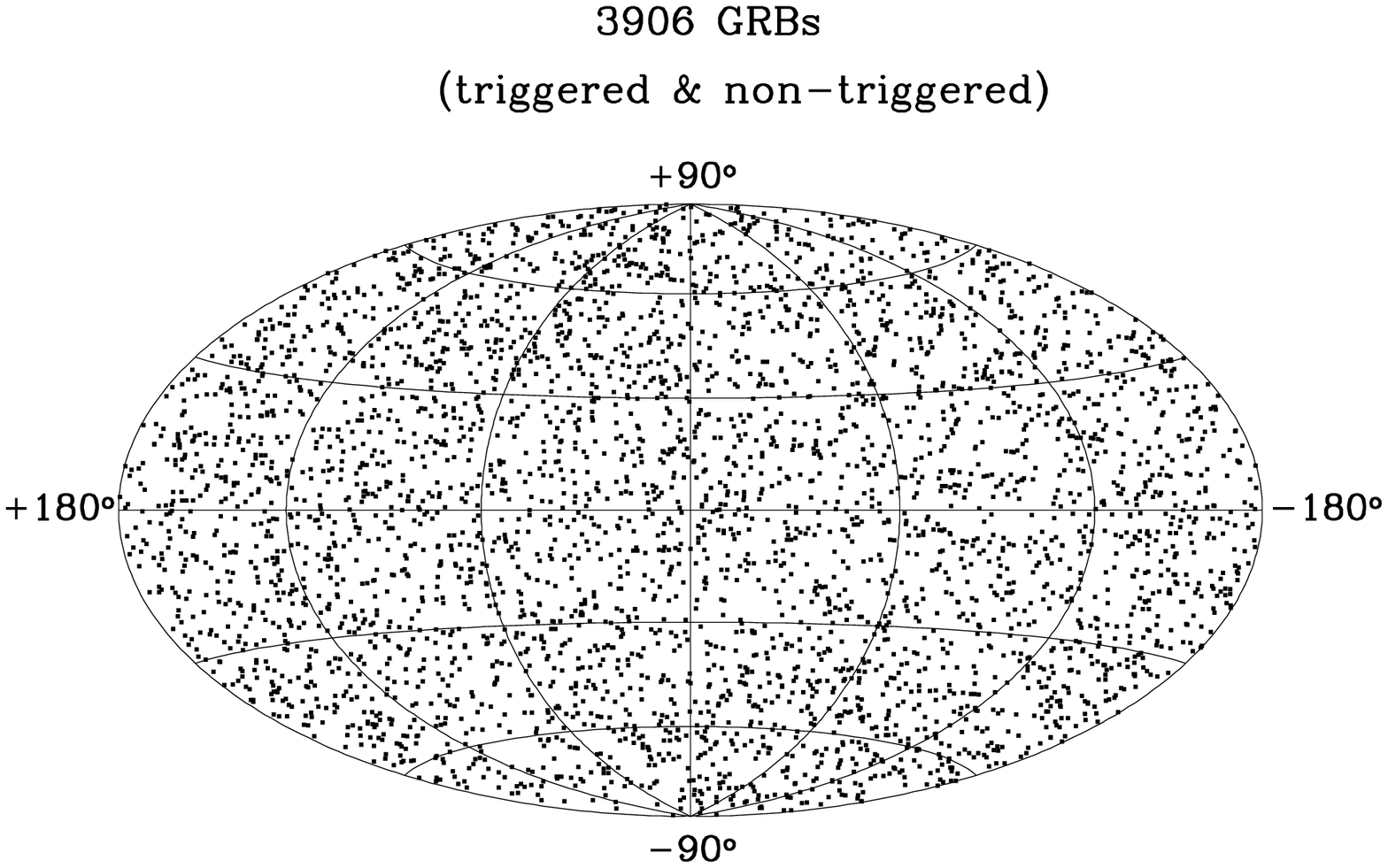}} }
\figcaption{
Sky distribution of all 3906 GRBs  
on an Aitoff-Hammer projection in Galactic coordinates.
\label{fig:11}
}

\subsection{Comparison with the BATSE and Kommers et al. Catalogs} 

During the period of our scan  BATSE  triggered  2704  GRBs.  Therefore  we
missed 636 of the BATSE  triggered  GRBs.  We estimated at a previous stage
of the scan (Stern et al.  2000b)  that $\sim$ 70\% of these GRBs were lost
due to gaps in the DISCLA  data,  $\sim$ 20\% were too short to be detected
at 1.024 s  resolution,  and $\sim$ 10\% were  missed due to various  human
mistakes.

K98 scanned the time interval TJD 8600--10800  and found 873  non-triggered
GRBs.  We found 1132 non-triggered bursts during that time interval, 745 of
them are in the catalog of K98.  Kommers (1999, private  communication; see
also K00)  inspected the 387 of our  non-triggered  events that K98 missed.
Kommers  confirmed 224 of these events as probable  GRBs.  24 of the events
were classified by him as particle  precipitations,  and 7 as noise.  90 of
our events were not classified as the off-line  trigger of K98 missed those
events due to nearby data gaps.  31 events were  classified as  ``unknown''
for their softness  (i.e., a very weak or no  significant  signal in energy
channel 3).

We checked  most of the 128 events from the K98 catalog that are missing in
our  sample.  We confirm  90 of them as GRBs, 13 we  classify  as  particle
precipitations, 4 as long delayed  aftercursors of triggered GRBs, and 2 as
solar  flares.  12 of these 128 events were  missed by our  trigger.  We do
not  include  the 90 GRBs of K98  confirmed  by us into our  sample  as the
efficiency we estimate does not account for such additional events.

The  comparison of our  estimates of the peak fluxes of GRBs in the 50--300
keV range with those of K98 and BATSE is shown in Figure~\ref{fig:12}.  One
can  see  a  reasonable  agreement  except  for  a  few  events  where  the
disagreement  could  result  from  errors  in the  setting  of the  fitting
windows.  This could, for  example,  cause part of an event to be  confused
with the  background.  In general, the agreement  between our and the BATSE
estimates  is  better  than  between  our and  those  of K98 in the  medium
brightness range where a meaningful comparison can be made.

\medskip 
\centerline{\epsfxsize=8.5cm{\epsfbox{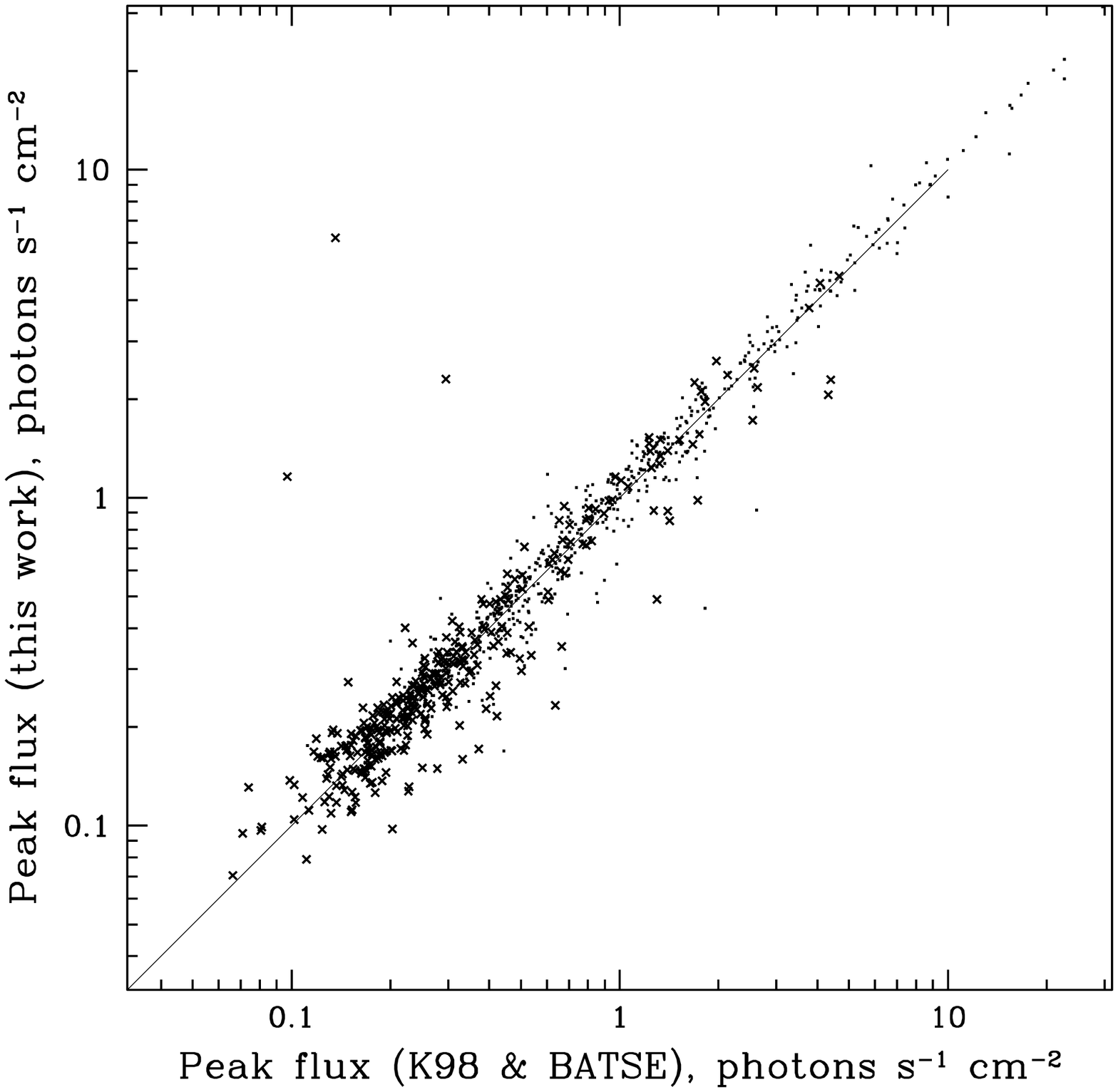}} }
\figcaption{
Our  estimates  of the peak flux in the  50--300  keV range  versus the K98
estimates  for the  same  non-triggered  bursts  (crosses)  and  the  BATSE
estimates  for  triggered  bursts  (dots) in 1.024 s time  resolution.  For
clarity, only first $\sim 30\%$ of the sample is shown.
\label{fig:12}
}
\medskip
\centerline{\epsfxsize=8.5cm {\epsfbox{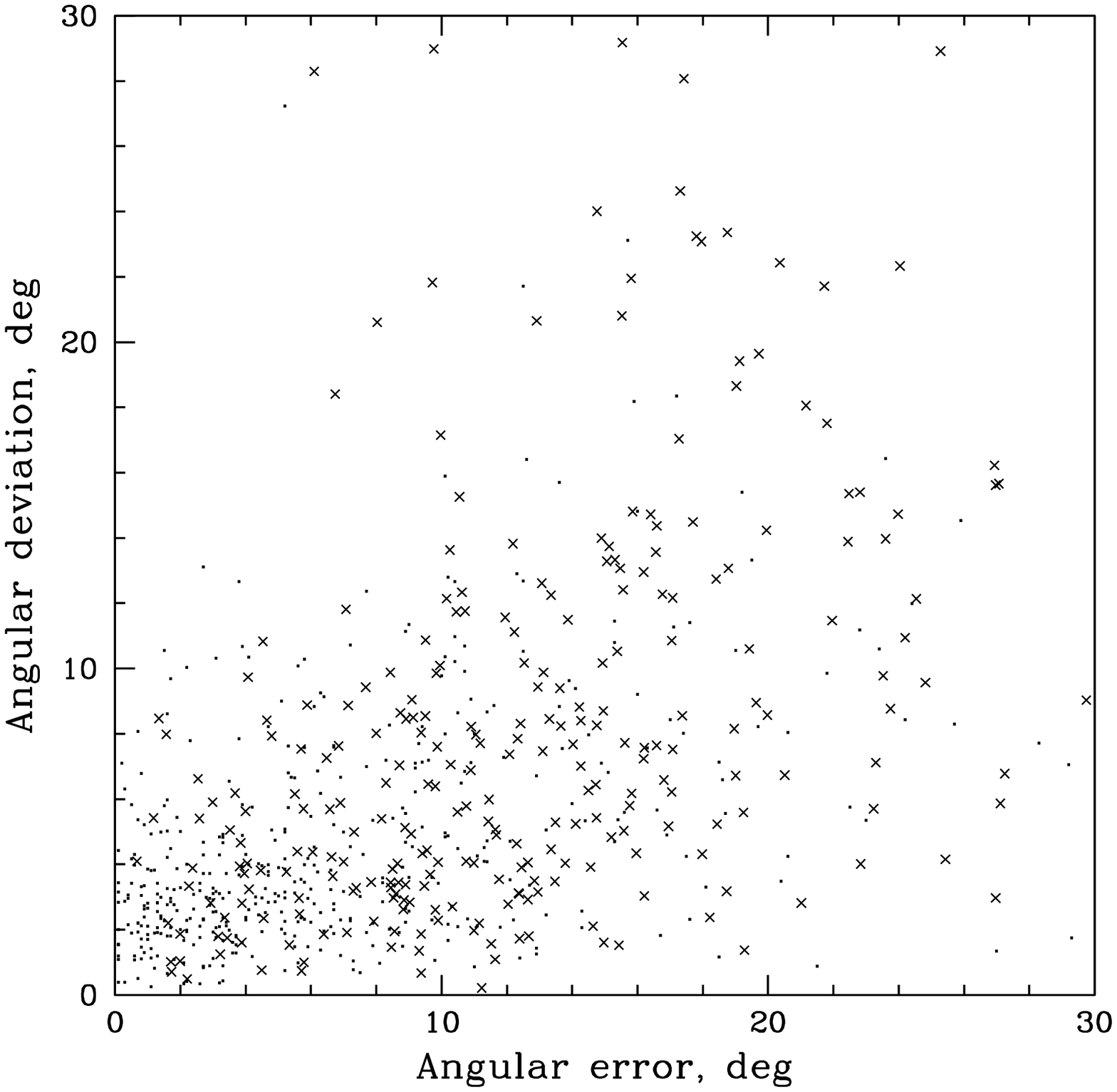}} }
\figcaption{
Angular deviations between our and the K98 best fit locations (crosses) and
between  our  and  the  BATSE   locations   (dots)   versus  the  $1\sigma$
localization  errors  according to our  estimates.  For clarity, only first
$\sim 30\%$ of the sample is shown.
\label{fig:13}
}
\medskip

The  deviations  between our best  locations and those from the K98 and the
BATSE  catalogs are shown in  Figure~\ref{fig:13}.  Most of the  deviations
are within the $1 \sigma$ errors.  There is, however, a substantial tail to
much larger deviations caused by systematic  errors.  Very large deviations
can result from the choice of different  local minima in the location  fit.
The peak in the  distribution  of the deviations  between our and the BATSE
locations  is  between  2 and 2.5  degrees  for  bright  GRBs.  This  value
approximately characterizes our typical systematic error for the locations.
The average  systematic error of LOCBURST as estimated by P99 is $\sim 1.7$
degrees.  Our procedure is not optimized for precise localization of strong
bursts, since we do not take into account  nonlinear  effects  occurring at
very high count rates and do not turn off less illuminated detectors in the
fit as LOCBURST does.  Our scheme is adjusted for  convenient  and reliable
fits of weak bursts.

The  uncertainties in the  interpolation of the background is an additional
source of systematic  errors which can be substantial  for weak GRBs.  This
bias is difficult to estimate.  It can be responsible  for large (more than
$1 \sigma$) deviations for weak events.

\medskip
\centerline{\epsfxsize=8.5cm {\epsfbox{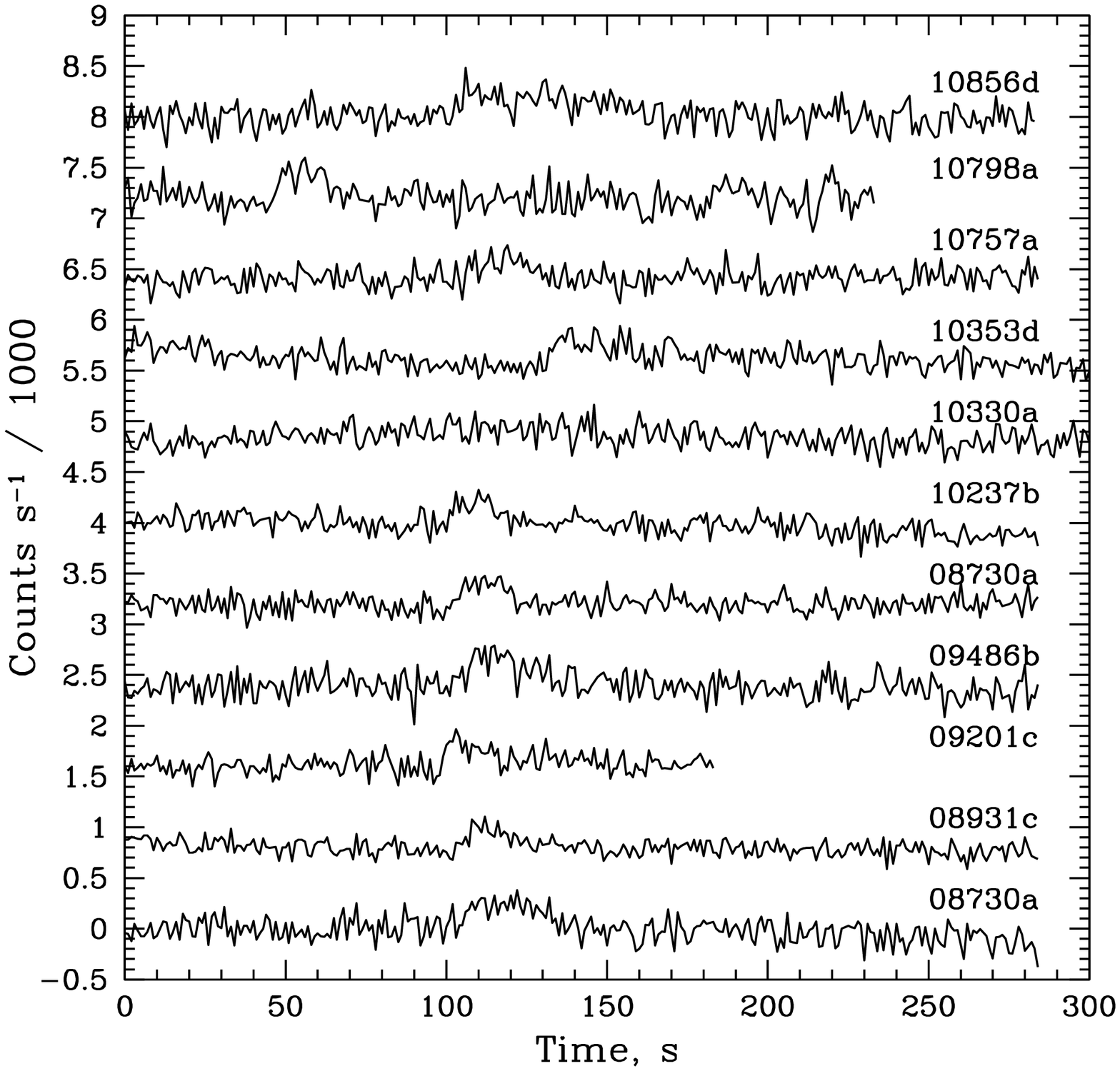}} }
\figcaption{
Fitting count rate time  profiles for the eleven  weakest GRBs  detected in
our scan.  The burst 10330a can hardly be  recognized  on this scale due to
its long duration ($\sim 150$ s) and  smoothness.  Nevertheless,  this is a
confident event clearly  visible on a longer time scale and at a wider time
binning.  The time profiles have been shifted vertically for clarity.
\label{fig:14}
}

\footnotesize
\begin{center}
{\sc TABLE 1\\
Parameters for the 11 weakest GRBs in the sample}
\vskip 4pt
\begin{tabular}{cccrcc} 
\hline
\hline
 ID             & $c$  &  $I$  &
 Sign. & $\delta_1$  & $\delta_4$  \\
 & & & & (deg) & (deg) \\
\hline
08730a & 0.049 & 0.072 & 17  & 18 & 37 \\
08913c & 0.049 & 0.088 & 11  & 23 & 43 \\
09201c & 0.045 & 0.076 & 15  & 33 & 70 \\
09486b & 0.045 & 0.076 & 14  & 24 & 47 \\
09882b & 0.047 & 0.077 & 9   & 33 & 61 \\
10237b & 0.047 & 0.081 & 10  & 39 & 84 \\
10330a & 0.045 & 0.072 & 8.5 & 37 & 67 \\  
10353d & 0.038 & 0.066 & 16  & 20 & 41 \\
10757a & 0.047 & 0.081 & 7.5 & 24 & 75 \\
10798a & 0.044 & 0.075 & 9.5 & 39 & 84 \\
10856d & 0.038 & 0.088 & 14  & 44 & 60 \\
\hline
\end{tabular}
\end{center}
\setcounter{table}{1}
{
ID is the event identifier consisting of TJD  plus an identifying letter;
$c$ is peak count rate in
units of  $\counts\secinv\cminvsq$;  $I$ is peak  photon  flux in  units of
$\ph\secinv\cminvsq$; significance is estimated using the residual $\chi^2$
relative to the linear  background in units of $\sigma = \sqrt {2N}$, where
$N$ is the number of the degrees of freedom;  $\delta_1$ and $\delta_4$ are
the sizes of the $1\sigma$ and $4\sigma$ confidence areas (see \S 4).
}
\medskip
\normalsize
\subsection{Bursts near the Detection Threshold} 

The  weakest  events of the sample  have peak  fluxes  below $0.1  \ph
\secinv  \cminvsq$.  They have been  detected  due to their  long  duration
using the 4 or 8 second trigger time scales.  They are not necessarily  the
least significant events.  One can get a feeling of the weakest part of our
sample from Figure~\ref{fig:14} and Table~1,  where we present
information on the 11 weakest events detected in the scan and classified as
GRBs.  The  events  in  Figure~\ref{fig:14}  and  Table~1  are
marked by their names consisting of the TJD when the event was detected and
an identifying letter.  All are more than 20 s long (shorter events are not
significant at these brightnesses), are significant, and have good $\chi^2$
maps.  The detection efficiency in this brightness range is $\sim 0.05$.

\subsection{The Data Archive} 

The electronic archive\footnote[8]{The    archive    is    available    at:
http://www.astro.su.se/groups/head/grb\_archive.html} (Stern  \&
Tikhomirova 1999) contains
the  following  data:  

\begin{enumerate}  
\item Table with event identifiers, arrival times, coordinates, peak fluxes,
and duration estimates ($T_{90}$ and $N_{50}$, i.e.,
the number of 1.024 s bins in original DISCLA data,
where the signal exceeds 50\% of the peak value).
BATSE trigger numbers are given for  triggered  events.
Events  identified  with the GRBs of the K98 catalog are marked.
Events in proximity  of the data gaps are also marked.
See Table~2.

\item Fragments of raw DISCLA data in FITS format covering the events.  The
typical time  interval is 300 seconds for shorter  ($<100$ s) events.  For
longer events the time interval is extended.

\item  Time  profiles  as the sum of the count  rates in the two  brightest
detectors  and in energy  channels  \#2 and \#3 (the  50--300  keV band) in
ASCII format.  This is the simplest and the most compact representation for
the time profiles.

\item Time  profiles of GRBs as the  fitting  count  rate,  $C_{ik}$,  with
subtracted  background,  see \S 3 and \S  5.1,  for  the  four  LAD  energy
channels in ASCII format.  This  representation  has a better  signal/noise
ratio than the sum of the count rates in any combination of detectors.

\item All distributions presented in this paper in both graphical and 
numerical form. 
\end{enumerate}

\section{Tests for non-GRB Contaminations} 

All our tests for non-GRB  contaminations  were  performed on the sample of
3678 events detected up to TJD 11499 (8.6 years of observations).

\subsection{Possible Types of the Contamination} 

We can divide the possible non-GRB contaminations of the sample into the
following categories:
 
1.  Poisson or non-Poisson background  fluctuations.  The least significant
GRBs of our sample are too  significant  to be  reproduced  by pure Poisson
fluctuations.  There are however  significant  variations of the background
due to ionospheric  phenomena,  occultations of astrophysical  sources, and
noise-generating   sources.  Such   events   do   not   pass   either   the
$\delta_4$-criterion   or  the   ``isolation''   criterion   (see   \S  4).
Nevertheless,  we  should  admit  the  existence  of  some  cases  where  a
fluctuation  may pass  all  tests.  This  kind of  fluctuations  is  almost
time-symmetric.  Therefore we can test for such a kind of contamination  by
searching for  ``negative  bursts'' in  sign-inverted  count rate data.  We
performed  such   tests  for 200 days of data  records  taken in  different
intervals over a few years representing various background conditions.  The
rate of triggers was practically  the same as in the normal scan.  However,
only 2 events passed the objective criteria  described in \S 4 (mainly they
were  rejected by the  $\delta_4$  criterion).  Both  events are within the
''$2\sigma$  area'' of Cyg X-1 and both are not  enough  isolated,  so they
should be rejected  subjectively.  Thus we constrain the  contamination  of
sign-symmetric  background fluctuations to be less than $\sim 0.5$\% of the
events.

2.  Ionospheric   events   such   as   particle   precipitation.   Particle
precipitation is a frequent phenomenon taking place at high latitudes:  the
satellite   flies   through  a  cloud  of  energetic   particles   emitting
bremsstrahlung.  Then a more or  less  time-symmetric  hump  of  counts  is
recorded  in  all  or  most  of  the  detectors.  Such  events  are  easily
recognizable by comparable signals in detectors facing opposite  directions
and,  correspondingly, by a bad $\chi^2$ map (many local minima of $\chi^2$
over a wide area of the sky) and a large  residual  $\chi^2$  (above 0.1 of
its initial value).  If the precipitation  occurs at some distance from the
satellite  path, the time profiles in different  detectors do not match and
again we obtain a bad location fit.  The most dangerous  events are distant
ionospheric  flares with fast (shorter than hundred  seconds)  variability.
We observed such events  recognizing them by ``context'',  i.e., they occur
at a high general ionospheric activity at high latitudes and appear in long
series.

3.  Solar flares and known  variable  X-ray  sources (Cyg X-1, bright X-ray
transients, and known X-ray  pulsars).  Solar flares are not a considerable
source of the  contamination  as they are much softer  than GRBs.  There is
some small  overlap in  hardness.  For this  reason, we reject  some events
which could be soft GRBs if they are consistent in location with the Sun.


4.  Unknown  astrophysical  sources emitting bursts similar to GRBs.  These
could be  Galactic  accreting  black  holes,  hard  bursts  of other  X-ray
pulsars,  or unknown  phenomena  (i.e.,  huge  stellar  flares).  We cannot
discriminate  such events  individually  if they overlap with GRBs in their
characteristics.  We can  only  sense  such  contaminations  statistically.
Such a statistical analysis is presented below.

\subsection{Hardness Ratios} 

GRBs are on average  harder than any other kind of events  that can mimic a
GRB.  We plot several  examples of non-GRB events of known origin  together
with GRBs in a hardness -- brightness diagram,  Figure~\ref{fig:15},  where
the  hardness  ratio is that  between the counts in the 50--300 keV and the
20--50  keV  energy  bands.  Outbursts  of Cyg X-1 have an almost  constant
hardness  ratio.  It is slightly  smaller than the average  ratio for GRBs.
Ionospheric  flares have a wider dispersion and partly overlap with GRBs in
their  hardness  ratios,  while being  softer on  average.  Bursts of X-ray
pulsars  are on average  considerably  softer  than  GRBs.  However,  their
hardness   distribution   overlaps   with   that  of  GRBs  and  some  hard
well-isolated  pulses  could be  included  into the  sample as GRBs.  Large
(tens of  events)  contamination  of this  kind  would be  revealed  in the
angular  distributions  of GRBs.  The softest events we deal with are solar
flares.  Their  overlap  with  GRBs in  hardness  is very  small as seen in
Figure~\ref{fig:15}.

If the weakest part of our sample were considerably  contaminated with some
non-GRB events, we could observe this in the hardness  distribution of weak
events  (this is hardly so for Cyg X-1 which has almost  the same  hardness
ratio as weak GRBs).  However, the hardness  ratios for the weakest  events
smoothly extend the hardness versus brightness correlation of brighter GRBs
as seen in  Figures~\ref{fig:15}  and \ref{fig:16}.  Thus the statistics of
the  hardness  ratios  for the  weakest  bursts  does not  demonstrate  any
indications of a considerable contamination.

\medskip
\centerline{\epsfxsize=8.5cm {\epsfbox{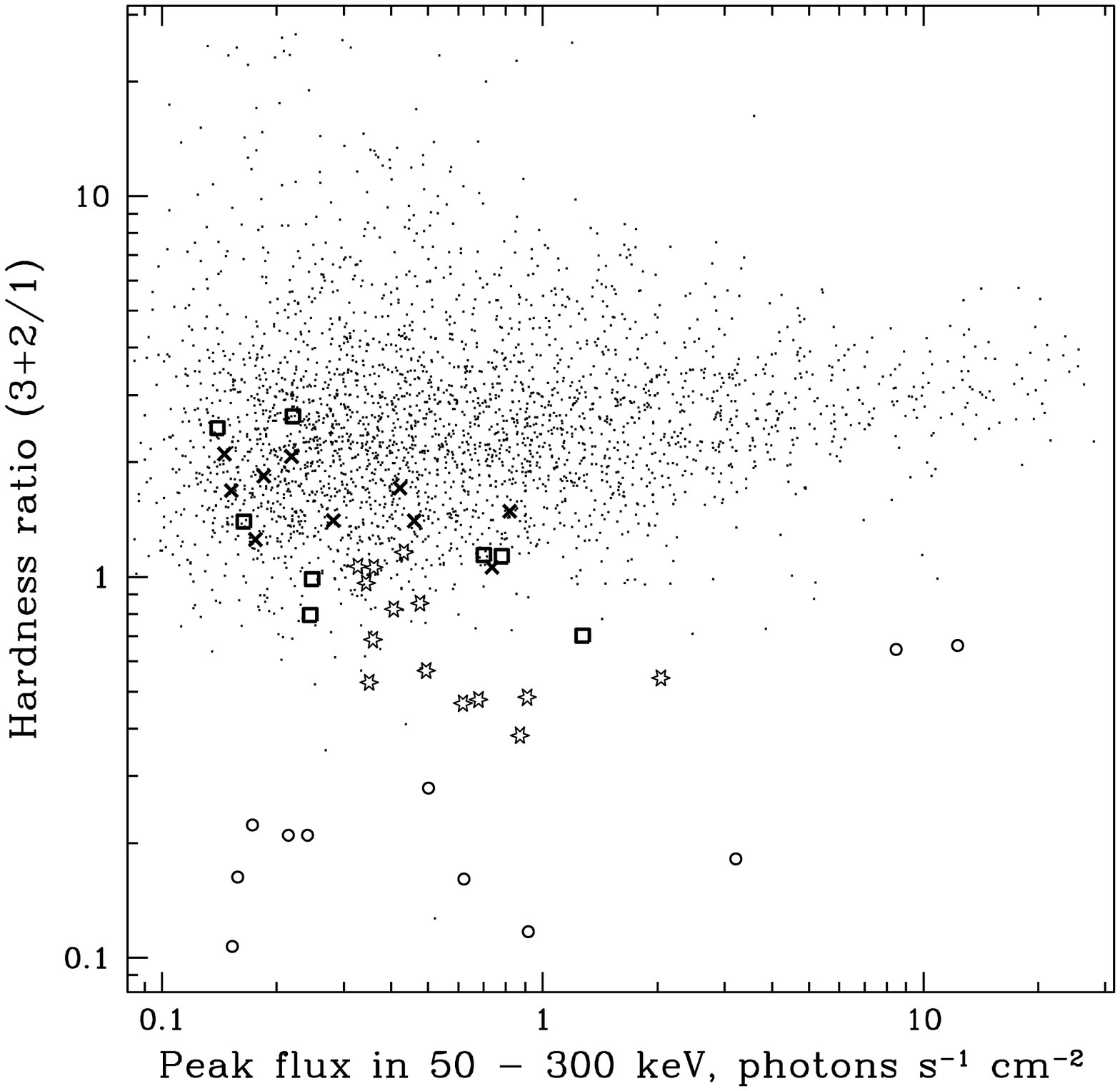}} }
\figcaption{
Hardness ratio (the integral of the counts in channels \#2 and \#3, 50--300
keV, to that in channel 1, 20--50 keV) versus the 50--300 keV peak flux for
GRBs (dots) and for  different  kinds of non-GRB  events found in the BATSE
records.  Crosses:  Cyg X-1; squares:  ionospheric phenomena; stars:  X-ray
pulsars (two objects), circles:  solar flares.
\label{fig:15}
}  
\medskip

\centerline{\epsfxsize=8.5cm {\epsfbox{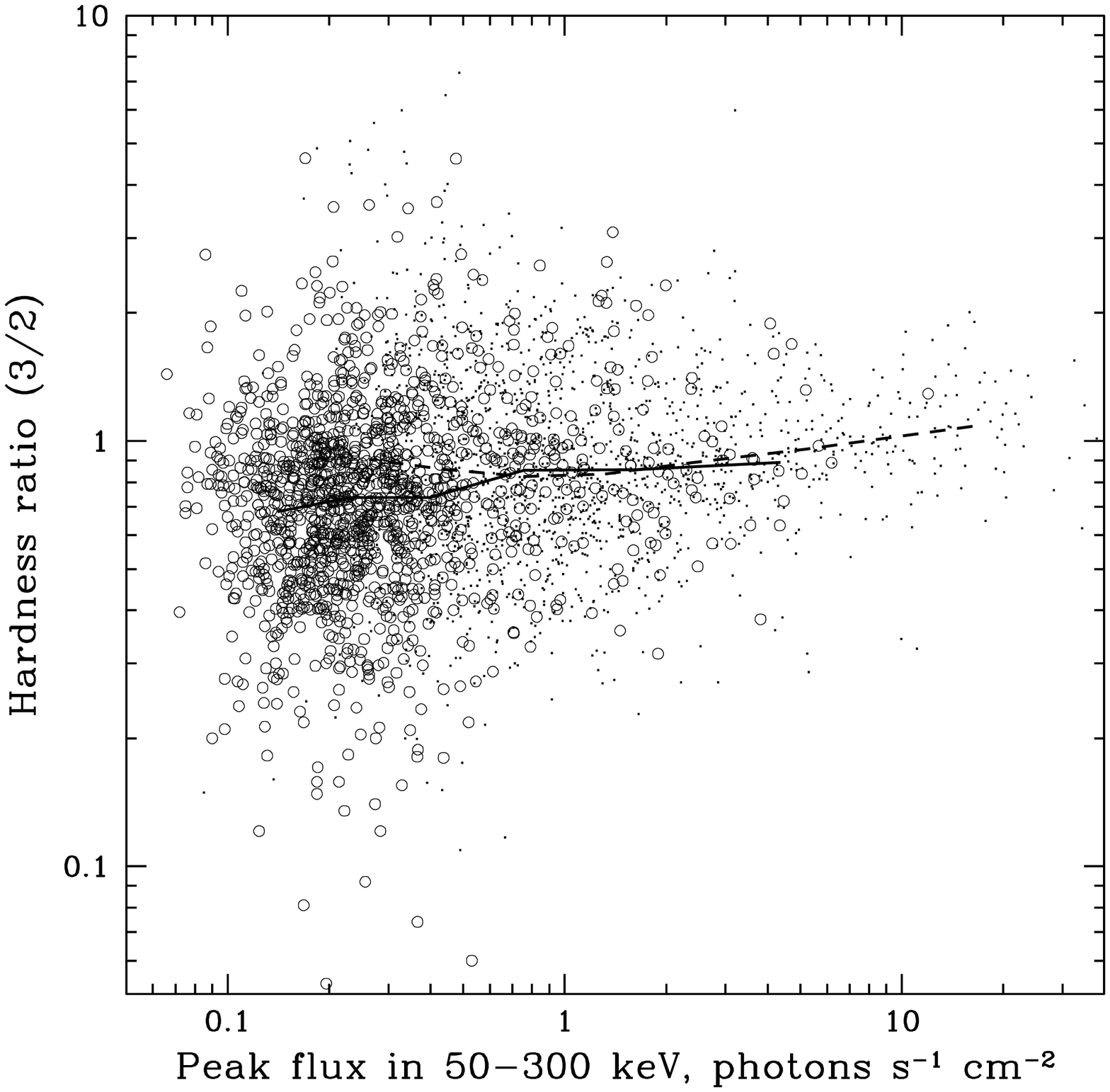}} }
\figcaption{
Hardness  versus the  50--300  keV peak flux for the GRBs  detected  in the
scan.  The  hardness  ratio is here the ratio of the  time-integrated  peak
count rate in the BATSE LAD channel \#3  (100--300  keV) to that in channel
\#2  (50--100  keV).  Solid  curve  shows the  median  hardness  ratio  for
triggered  GRBs, while the dashed  curve  shows the same for  non-triggered
GRBs.  Circles represent the 1838 non-triggered  GRBs, while dots represent
the 2068  BATSE-triggered  GRBs detected in the scan.  Weak  triggered GRBs
are harder than  non-triggered  ones in the same peak flux range because of
short  hard  events.  They are weak at 1024 ms  resolution  but   are
easily  triggered at 64 ms  resolution.  The general  trends  approximately
agree with the results of Nemiroff et al.  (1994).
\label{fig:16}
}
\medskip

\subsection{Angular Distributions} 

Each  kind  of  non-GRB  events  has  its  specific  non-isotropic  angular
distribution  which can affect the  angular  distribution  of events of the
sample  if the  latter  is  contaminated  with  non-GRBs.  Here we  analyze
several  angular  distributions  of events in the  sample.  Each of them is
sensitive to some kind of non-GRB contamination.  Since the weakest part of
the sample is more  likely to be  contaminated,  we  analyze  some  angular
distributions  using the subsample of the weakest  GRBs.  The polar angular
distributions  for both real and test bursts are consistent  with the BATSE
sky   exposure    function    (Paciesas   et   al.   1999)   as   seen   in
Figure~\ref{fig:17}.

\medskip
\centerline{\epsfxsize=8.5cm {\epsfbox{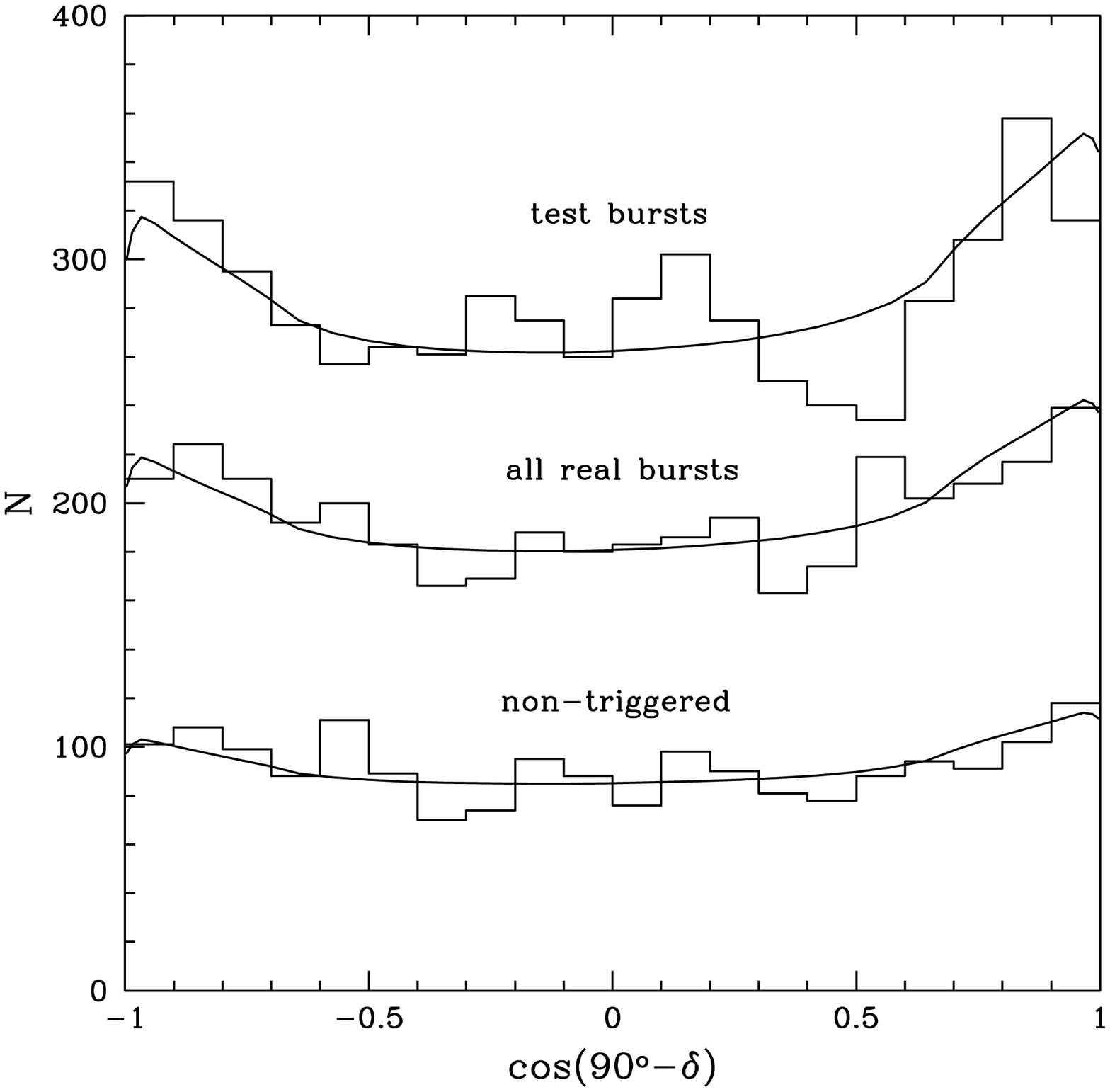}} }
\figcaption{
Polar angular  distributions  of GRBs and test bursts.  Smooth  curves show
the  BATSE  sky  exposure   function   (Paciesas  et  al.  1999).  The  GRB
declination is denoted by $\delta$.
\label{fig:17}
}
\medskip
\centerline{\epsfxsize=8.5cm {\epsfbox{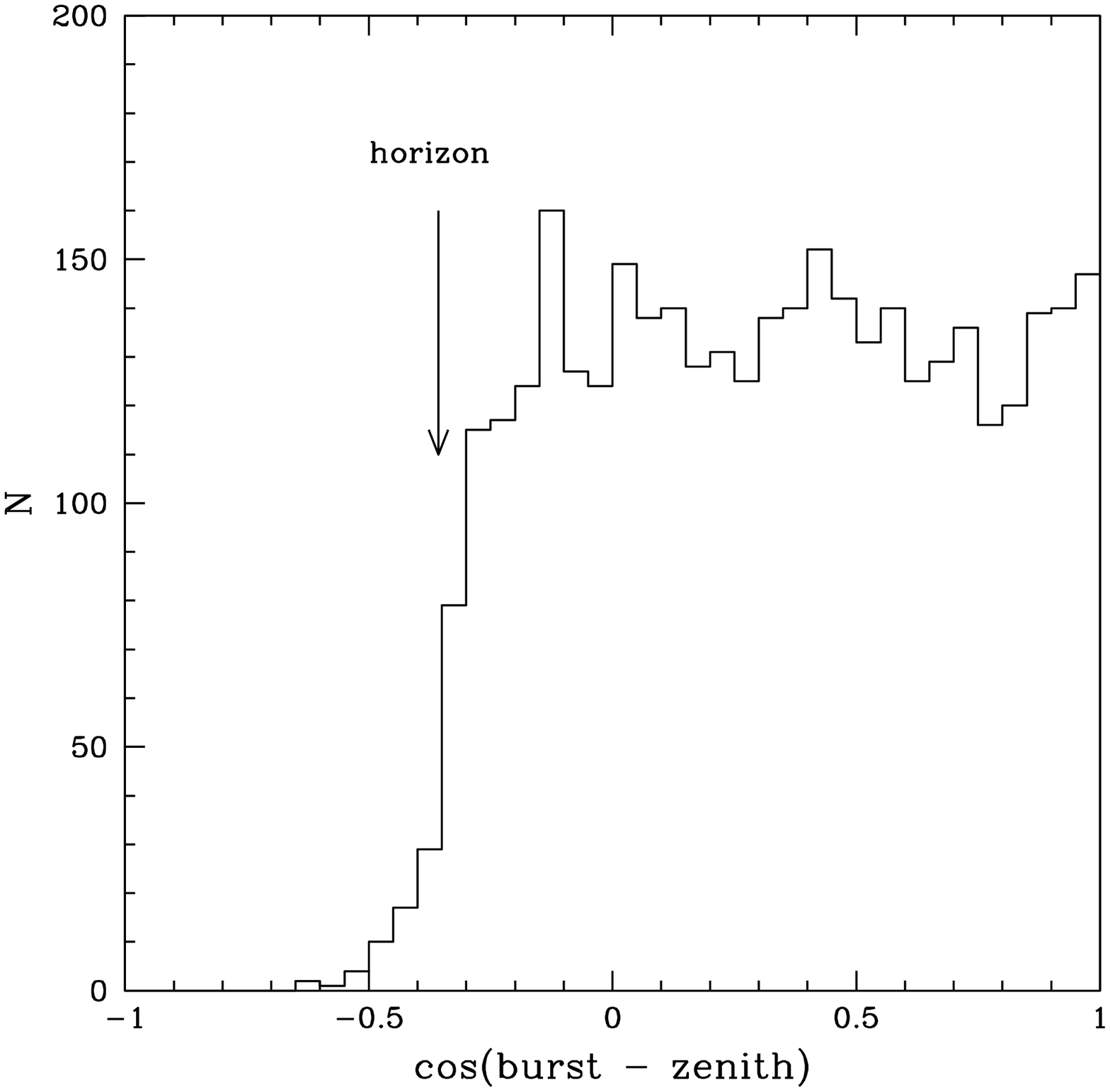}} }
\figcaption{ 
Distribution  of locations of GRBs relative to zenith.  The arrow shows the
mean  position of the Earth's  horizon.  (The cosine of the  horizon  angle
varied  between about  $-0.31$ and $-0.38$ due to the changing  altitude of
{\it CGRO}.)
\label{fig:18}
}
\medskip

The  distribution  of GRBs  relative  to the  Earth's  horizon  is shown in
Figure~\ref{fig:18}.  The  shape  of the  distribution  is  reasonable:  an
isotropic  distribution  above the horizon and a clear step at the horizon.
There is no excess of events  towards the horizon which could appear if the
sample  contains many  misidentified  events of  terrestrial  (ionospheric)
origin.

No excess of events  towards  the Sun has been  found.  Figure~\ref{fig:19}
demonstrates a very wide separation between solar flares and GRBs in a plot
of hardness ratio versus angular  distance from Sun.  One can conclude from
Figure~\ref{fig:19} that the contamination of the sample by solar flares is
negligible.

\medskip
\centerline{\epsfxsize=8.5cm {\epsfbox{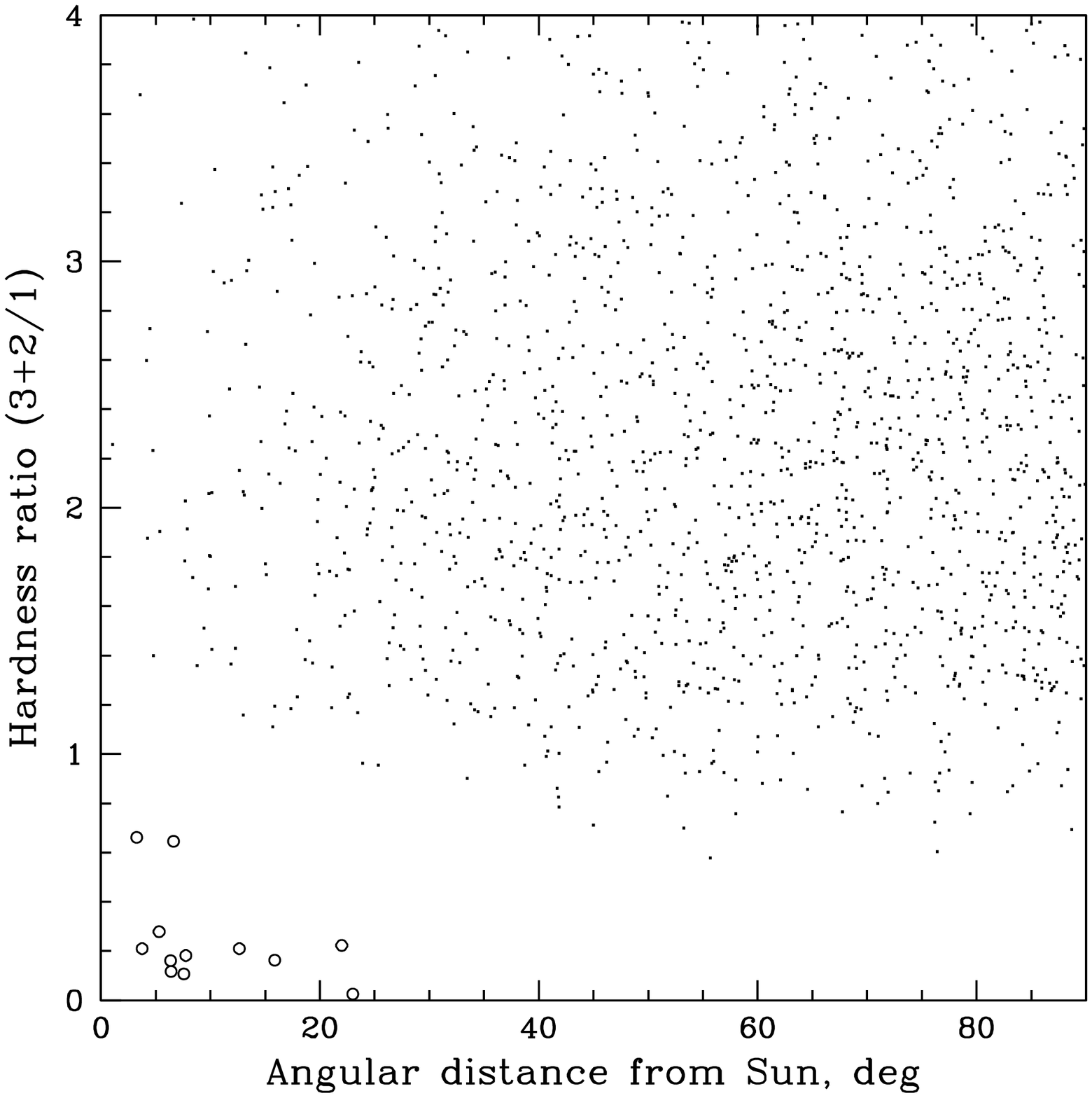}} }
\figcaption{Hardness ratio vs. angular separation from the Sun for 
GRBs (dots) and several solar flares (circles).
\label{fig:19}
}
\medskip
\centerline{\epsfxsize=8.5cm {\epsfbox{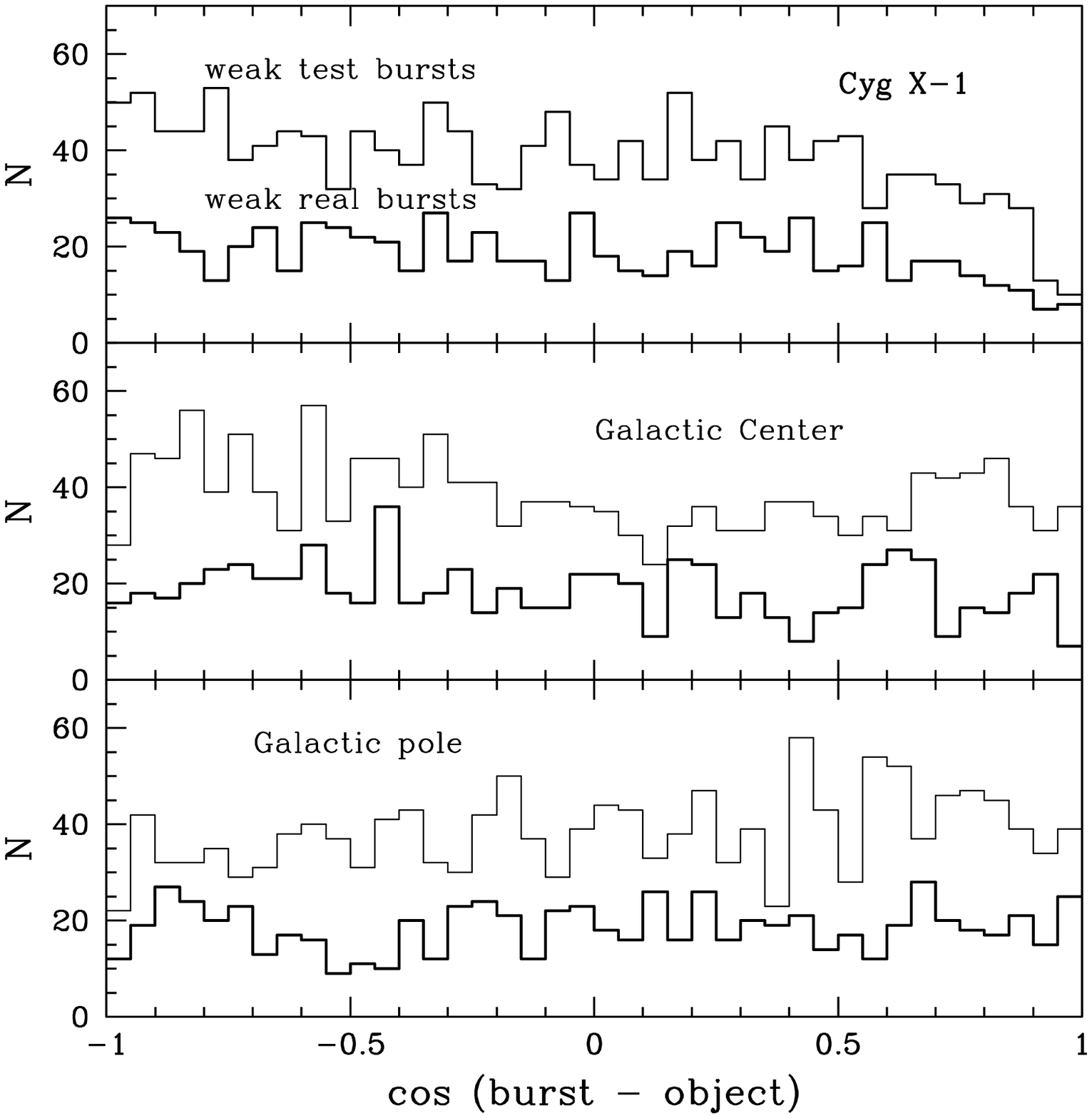}} }
\figcaption{
Angular  distributions  of  detected  weak  ($P < 0.2  \ph\secinv\cminvsq$)
bursts  relative to Cyg X-1, the Galactic  Center, and the  Galactic  pole.
Thin line histograms are for weak test bursts, and thick line histograms --
for weak real bursts.
\label{fig:20}
}
\medskip

The  distribution  of weak GRBs and weak test  bursts ($P < 0.2 \ph \secinv
\cminvsq$)   relative   to  the   direction   to   Cyg~X-1   is   shown  in
Figure~\ref{fig:20}.  There is an  evident  deficit  of both  test and real
bursts in the direction to Cyg~X-1.  This is natural as we missed many weak
bursts in that  direction  when  subtracting  the signal from Cyg~X-1.  The
depression   is   slightly   deeper  in  the  case  of  test   bursts.  The
corresponding  expected  number of real bursts in the two last bins (a cone
of half opening  angle  $26^{\rm  o}$ around Cyg X-1) is 11, while the real
number is 15.  The  difference is not  significant,  of course.  The formal
$1\sigma$  confidence  interval  for  contamination  by Cyg~X-1  is $4\pm5$
events.  Brighter outbursts of Cyg~X-1 are very rare.

Contaminating X-ray pulsars and variable X-ray sources would give an excess
of weak bursts in the direction of the Galactic Center.  Actually there is,
on the  contrary,  a  deficit  of both  real and test  weak  bursts  in the
direction  of the  Galactic  Center.  A small  fraction of this  deficit is
explained by the coverage  function.  A stronger effect appears as a result
of worse  background  conditions in that  direction,  mainly due to Cyg X-1
($\sim 60^\grad$ from the Galactic Center direction).

The  formal   $1\sigma$   confidence   interval   for  a  Galactic   Center
subpopulation  in our sample for a cone of $37^\grad$ half opening  angle
in the direction of the Galactic  Center  (where both of the X-ray  pulsars
shown in Figure~\ref{fig:15} are located) is $-10\pm 10$ events.  Similarly
we do not see any excess of weak events towards the galactic plane.


\medskip
\centerline{\epsfxsize=8.5cm {\epsfbox{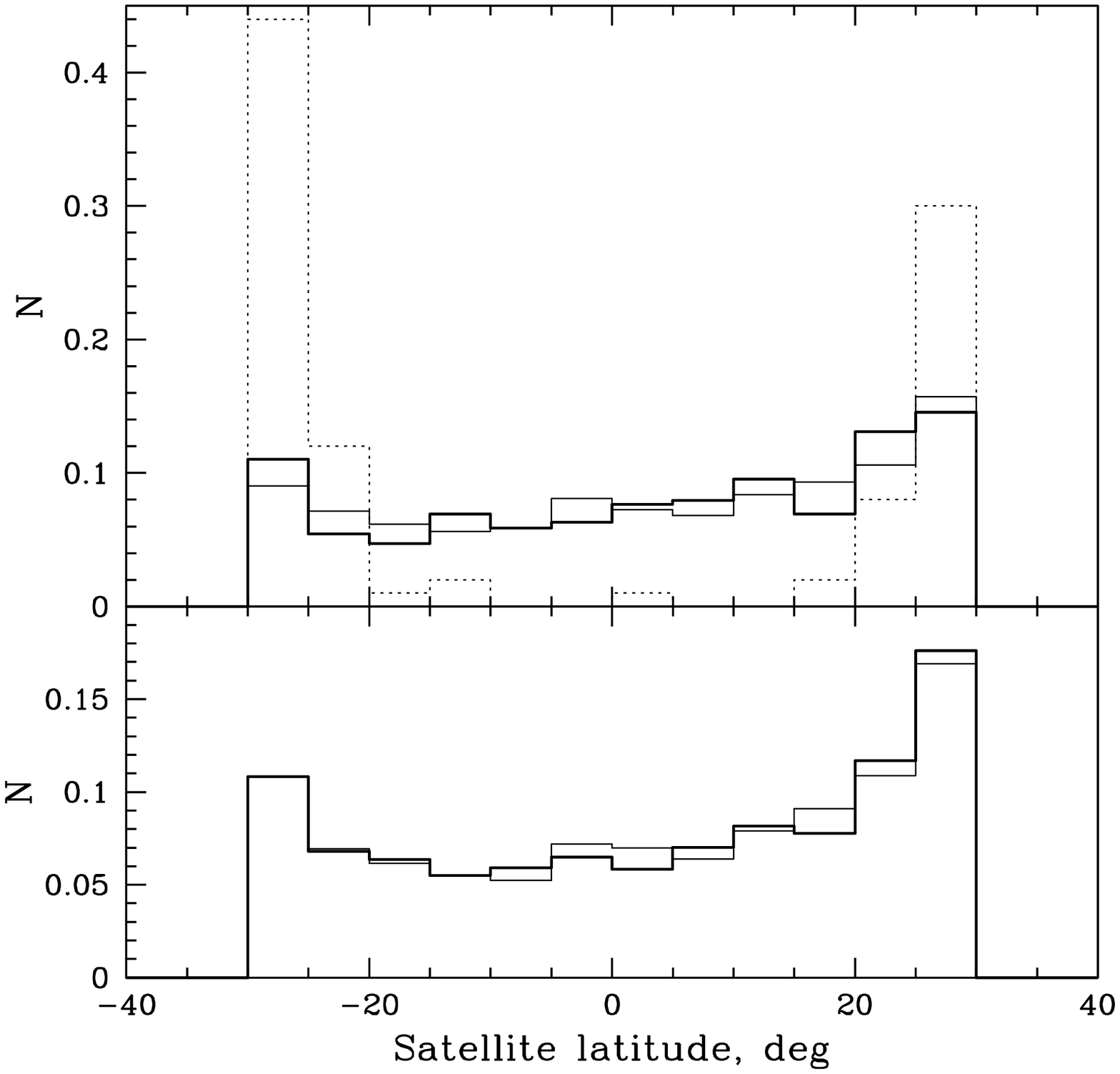}} }
\figcaption{
Normalized  distributions  of detected  bursts and  particle  precipitation
events  over the  latitude  of {\it CGRO} at the time of  detection.  Upper
panel:  weak  events  ($P<  0.2   \ph\secinv\cminvsq$);  lower  panel:  all
events.   Thick line histograms:  real bursts; thin line  histograms:  test
bursts;  dotted  line  histogram:  a sample of 100  particle  precipitation
events  detected  during  different  time  intervals of  observations.  The
asymmetric  shape  of the  burst  histograms  is  mainly  due to the  South
Atlantic  magnetic  anomaly where any detection is impossible  because of a
very large ionospheric background.
\label{fig:21}
}

\subsection{Dependence on the Satellite Latitude}  

 The intensity of ionospheric  activity  strongly  depends on the satellite
latitude.  Particularly, the frequency of particle  precipitations  sharply
increases at high latitudes.  Therefore the  distribution  of the satellite
latitude,  when an event of our sample was  recorded, is  sensitive  to the
contamination of the sample by the ionospheric  events.  Then we would have
an excess of events detected at large Northen and Southern  latitudes.  The
distribution  of events along the  satellite  path should not be uniform as
the background  conditions are substantially  worse at a high latitude.  As
far as we have common conditions for the detection of real and test bursts,
an excess of real  events  relative  to the test  bursts can be  estimated.
Figure~\ref{fig:21}  shows the  results  of such a test for all and for the
weakest  events.  There is no excess of real  events  relative  to the test
bursts at high  latitudes.  An  approximate  $1\sigma$  upper limit for the
contamination  of the sample by events of  ionospheric  origin at latitudes
above $\pm 20^\grad$ (where more than 90\% of the particle  precipitation
events occur, see  Fig.~\ref{fig:21})  is 42 events  (i.e.,  1.1\%) for the
full sample and 17 events  taking into account only the weakest part of the
sample where a misclassification  is more probable (at $P < 0.2 \ph \secinv
\cminvsq$, the 680 weakest GRBs).

\subsection{Summary of Contamination Tests}

 Summarizing the results of the tests:
\begin{enumerate}
\item
 The contamination of the sample by solar flares is negligible.
\item
 The Cyg X-1 contribution is estimated as $4\pm5$ events. 
\item
 The contribution of all sources concentrated within $37^\grad$ around the 
Galactic Center direction is constrained to be $-10\pm 10$ events. 
\item
The least  constrained  source of contamination is produced by ionospheric
phenomena because the  corresponding  angular  distributions  are the least
determined.  However, even in this case the contamination is constrained to
be at the level of $\sim 1\%$.

\end{enumerate}

We should admit the  existence  of some  non-GRB  isotropic  background  of
events of a different nature but looking like GRBs.  Then this is rather an
issue of the  classification  of GRBs,  which is beyond  the  scope of this
paper.

\section{The $\log N - \log P$ Distribution} 

\subsection{The Efficiency Function and the Corrected $\log N - \log P$
Distribution} 
 
The efficiency function of the search is defined as the ratio of the number
of detected  test bursts to the number of test  bursts  applied to the data
versus  the {\it  expected}  peak  count  rate,  $c_e$.  It is shown by the
histogram in  Figure~\ref{fig:22}.  For many  applications, it is useful to
have a simple  analytical  representation  of the efficiency  function.  We
find that this function can be well approximated by
\be \label{eq:efffun}
E(c_e)= 0.70 \left\{1-\exp\left[-\left(\frac{c_e}{c_{e,\rmo}} 
\right)^2\right]\right\}^{\nu} ,  
\ee
where two parameters $c_{e,\rmo}=  0.097\counts\secinv\cminvsq$ and 
$\nu=2.34$ (see the smooth curve  in Fig.~\ref{fig:22})

The  result  of a  direct  correction  of  the  measured  peak  count  rate
distribution    (Fig.~\ref{fig:8}    and   the    dotted    histogram    in
Fig.~\ref{fig:23})    using   the   efficiency   function   is   shown   in
Figure~\ref{fig:23} by the solid histogram.  Such a correction is valid for
reconstructing the $\log N - \log P$ distribution only at peak brightnesses
that  sufficiently  exceed the  threshold  such that the  measured  and the
expected  count rates are almost  identical.  There, the errors in the peak
count rates are  moderate  and  symmetric  (see  Fig.~\ref{fig:6})  and the
efficiency changes smoothly.  One sees that the corrected $\log N - \log P$
distribution extends down to lower brightnesses almost straight without any
indication   of   a   turn-over.  This   result   differs   from   previous
interpretations  of the  data,  which  implied  that the  $\log N - \log P$
distribution  smoothly bends down at low brightnesses  (see the data points
of K00 in Fig.~\ref{fig:23}).
  
\medskip 
\centerline{\epsfxsize=8.5cm {\epsfbox{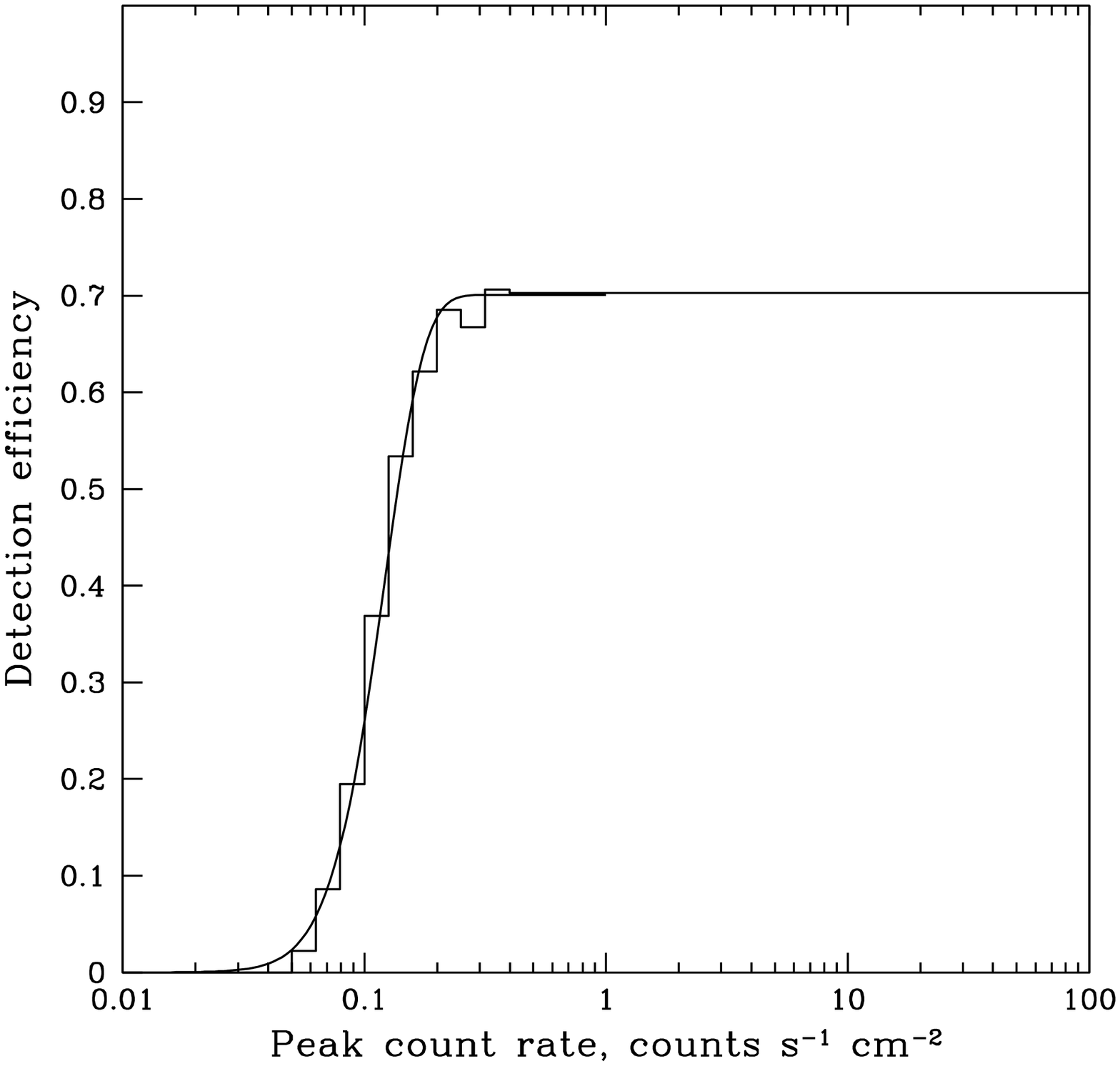}} }
\figcaption{ 
The efficiency of the off-line burst  detection  defined as the fraction of
the test bursts  detected in our scan versus the peak count rate, $P$.  The
efficiency is normalized to the total number of bursts  occurring above the
Earth's  horizon  (all test bursts were  sampled  above the  horizon).  The
smooth curve shows the fit given by equation~(\ref{eq:efffun}).
\label{fig:22}
}
\bigskip
\centerline{\epsfxsize=8.5cm{\epsfbox{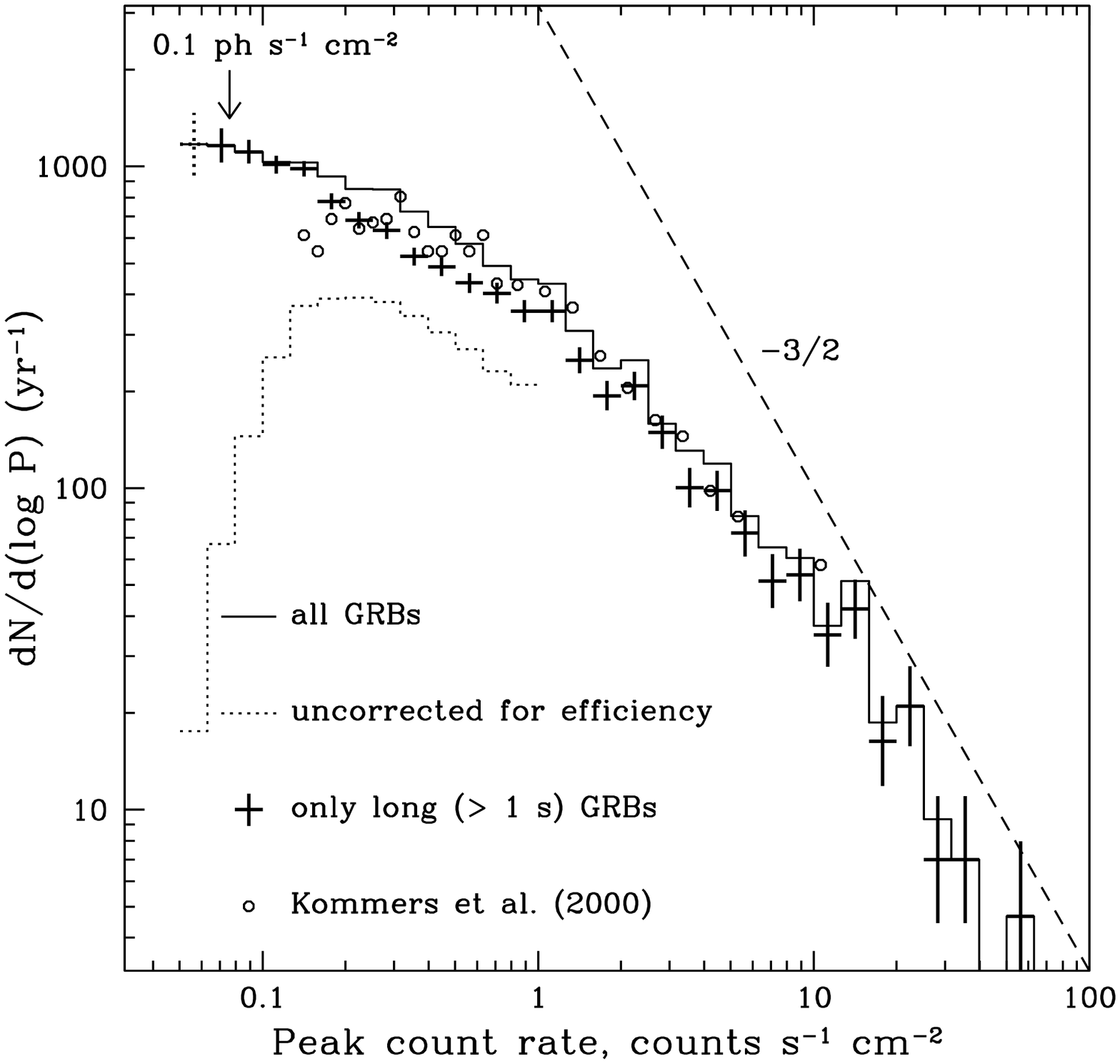}} }
\figcaption{  
The  differential  $\log  N  -  \log  P$  distribution  corrected  for  the
efficiency  function in absolute  units for all 3906 GRBs  detected  in our
scan (solid histogram).  The original uncorrected  distribution is shown by
dotted  histogram.  The corrected  distribution  for the 3300 events in our
sample  with  a  duration  longer  than  1  s  is  shown  by  crosses.  The
corresponding distribution for the 2265 GRBs (of any duration) found by K00
is shown by  circles  (the K00 data  were  transformed  to  count  rates by
applying the factor 0.75, see \S 5.2 and Figure~\ref{fig:7}).  The leftmost
dashed data point is probably  affected by the threshold  bias arising from
errors in the peak  count  rates.  
\label{fig:23}
}
\medskip

Our whole sample is not homogeneous  because it contains short events where
the peak flux estimate in 1 s time  resolution  is wrong.  To eliminate the
corresponding  bias we excluded from the distribution all events consisting
of one bin (i.e.,  where only one 1.024 s bin is above 0.5 of the peak flux
value).  The  corresponding  efficiency  function was calculated using only
those test bursts that produce events satisfying the above criterion.  Then
we have a more or less  homogeneous  estimate of the peak count rates.  The
$\log N - \log P$  distribution  of only long events is shown by crosses in
Figure~\ref{fig:23}.  Such a distribution is more relevant for cosmological
fits than the total distribution.

\subsection{The  Efficiency Matrix and  $\log N - \log P$ 
Close to the Detection Threshold} 
 
When  the  burst  peak  flux  is  close  to the  detection  threshold,  the
difference  between the expected and the  measured  peak count rates can be
significant.  Therefore,  in order  to  reconstruct  the  $\log N - \log P$
distribution  at low fluxes, we have to take into account the matrix of the
probabilities  of detecting a burst of a given  expected  count rate with a
given  measured   count  rate.  Using  test  bursts  from  our  sample,  we
constructed  such a matrix.  The ``low  brightness''  end of this matrix is
presented in Table~3  (see also  Fig.~\ref{fig:6}).  We can see
the  asymmetry  of the matrix near the  threshold  (i.e., the  off-diagonal
elements  are  larger  than the  diagonal  ones)  which  results  from  the
selection bias discussed in \S 5.2.

In spite of the fact that we had $11\ 100$ test bursts, the  statistics  at
the low brightness end of the efficiency  matrix is relatively poor and the
error there is large.  We can improve the  situation by  approximating  the
matrix by a smooth  function of the expected and the measured  count rates,
$c_e$ and $c_m$, and by extrapolating this function to smaller count rates.
We approximate the efficiency matrix with a factorized expression where the
efficiency function, the peak brightness error, and the selection bias were
fitted independently:

\be \label{eq:effmatr}
F(c_e,c_m)= E(c_e)  \frac{1}{\sigma\sqrt{\pi}}
\exp \left[ -\frac{\log^2(c_m/c_{m,\rmo})}{2\sigma^2} \right],
\ee
where  $E(c_e)$  is given by  equation  (\ref{eq:efffun}),  the  log-normal
factor  describes the relative error of the measured  count rate, $\sigma =
0.09  (0.08/c_e)^{1/2}$,  and the  selection  bias is crudely  expressed as
$c_{m,\rmo}=c_e+0.05\exp (-c_e/0.05)$.

As was mentioned above, the correction using the efficiency function is not
exact  especially  near the threshold.  One should  deconvolve the observed
peak  brightness  distribution   (Fig.~\ref{fig:8})  using  the  efficiency
matrix.  This is easier to do with a forward folding  method, i.e., fitting
the observed  distribution using a convolution of a hypothetical  $\log N -
\log P$  distribution  with the  efficiency  matrix.  We cannot  use ``data
points'' like those in  Figure~\ref{fig:23}  for the hypothetical $\log N -
\log P$  distribution  as the  forward  folding in this case  allows  large
fluctuations  between  neighboring  data points and the result is unstable.
One should use some smooth  function as a hypothesis for the $\log N - \log
P$ distribution, convolve it with the efficiency matrix  (\ref{eq:effmatr})
and  compare  with the data.  Such a fit is beyond the scope of the present
paper.

\section{Conclusions} 

With a careful  off-line scan of the daily  archival  data, BATSE becomes a
more sensitive instrument with better known characteristics.  The detection
threshold  changes  by a factor 2, from  $\sim 0.2  \ph\secinv\cminvsq$  to
$\sim 0.1 \ph\secinv\cminvsq$.  The total number of detected GRBs increases
by 70 per cent.  If we consider  only long  events  (short  GRBs could be a
different  phenomenon),  the gain is a factor of $\sim 2$.  The recognition
of the weakest GRBs is still  confident and the  contamination  with events
originating from trivial kinds of non-GRBs is small.

As a first  result, using  the new  sample  of GRBs  with  known  detection
efficiency, we pointed out that the $\log N - \log P$ distribution does not
show a turn-over at lower  brightnesses  implied by previous  studies.  One
simple  consequence  is that  the  estimate  of the  number  of GRBs in the
visible  Universe  should be increased.  Just the  ``visible''  part of the
$\log N - \log P$ distribution  implies  1200--1300 GRBs occurring per year
at peak fluxes exceeding $0.1\ph\secinv\cminvsq$  (previous versions of the
$\log N - \log P$ distribution implied $\sim 600$ GRBs per year above $0.18
\ph\secinv\cminvsq$).  A possible  extrapolation  of the new $\log N - \log
P$ distribution  to lower  brightnesses  would probably imply a much larger
rate of up to several thousands of GRBs per year.

The best  possible  efficiency  and scan  quality  have  still not yet been
achieved.  When we repeated the scan for a fraction of data  scanned  early
on, we found  13 new  GRBs  per 60  days.  This  means  that an  additional
considerable increase of the statistics of the weakest GRBs is possible (as
is an extension of the  statistics of the weakest test  bursts).  There are
obvious  possibilities for improving the test burst method as was discussed
in \S 2 and in Stern et al.  (2000b), for improving  the  approximation  of
the background, etc.  This means that a new scan is desirable.


This research made use of data obtained  through the HEASARC Online Service
provided by  NASA/GSFC.  We are grateful to R.  Preece who supplied us with
valuable  information  and  suggested  one of the used tests for the sample
contamination.  We are  grateful to J.  Kommers who  inspected a large part
of our sample and thus helped us to  eliminate  non-GRB  contamination.  We
thank   A.~Skassyrskaia,   A.~Skorbun,   E.~Stern,  V.~Kurt,   K.~Semenkov,
S.~Masolkin,  A.~Sergeev,  M.~Voronkov,  and F.~Ryde for  assistance.  This
work was supported by the Swedish  Natural  Science  Research  Council, the
Royal Swedish Academy of Science, the Wenner-Gren Foundation for Scientific
Research,  and the Swedish  Institute.  D.~K.  was  supported by RFBR grant
97-02-16975.



\footnotesize
\begin{center}
{\sc TABLE 2\\
GRB characteristics}
\vskip 4pt
\begin{tabular}{lrccrrrrrcc}
\hline
\hline
ID\tablenotemark{a} & Time\tablenotemark{b} & 
BATSE Trigger\tablenotemark{c} & Peak Flux\tablenotemark{d}
& R.A.\tablenotemark{e} & Decl.\tablenotemark{e} & 
$\delta_1$\tablenotemark{e} & $T_{90}$\tablenotemark{f}
& {$N_{50}$\tablenotemark{g}} & Kommers & {Gap}\\
{ } & {(s)} & {or 1 if not} & 
{($\ph \secinv\cminvsq$)} &  {(deg)} &  {(deg)} & {(deg)}
& {(s)} & {(bin)} & {Catalog\tablenotemark{h}} & 
{Flag\tablenotemark{i}} \\
\hline
08369d & 71474 & 107  &   0.373 &  220.0 &  10.5   &    6.1 &  207 & 37& 0& 0\\
08370b &  6141 &   1  &   0.248 &   68.4 &  $-$6.4 &    8.6 &   83 & 21& 0& 0\\
08370f & 71005 & 108  &   0.178 &  209.8 & $-$45.6 &   34.1 &    4 &  2& 0& 0\\
08371a &  2208 & 109  &   3.725 &   89.0 & $-$16.0 &    0.2 &   95 & 27& 0& 1\\
08371c & 20208 & 110  &   0.454 &  341.8 &  25.0   &    7.9 &  549 &  6& 0& 0\\
08372a &  5386 &   1  &   0.336 &  246.2 &  65.4   &    7.6 &    8 &  5& 0& 0\\
08372d & 80024 & 111  &   0.502 &   78.1 & $-$22.5 &    3.8 &   84 & 25& 0& 0\\
08373a & 14897 &   1  &   0.257 &   32.4 & $-$46.1 &   21.4 &   85 &  5& 0& 0\\
08373c & 32689 & 114  &   0.540 &   76.5 & $-$19.2 &   11.1 &  370 &  5& 0& 0\\
08375b & 11481 & 121  &   1.324 &  173.2 &     9.1 &    1.9 &   87 & 13& 0& 0\\
08375c & 21537 &   1  &   0.259 &  315.3 & $-$48.9 &   17.2 &   50 & 13& 0& 0\\
08376d & 61501 & 130  &   3.561 &  133.6 &     1.9 &    1.4 &  203 & 15& 0& 0\\
08377a & 30013 & 133  &   0.621 &  124.3 &  $-$1.9 &    6.1 &  174 & 22& 0& 0\\
08378b & 40667 & 138  &   0.316 &  126.0 & $-$21.4 &   17.1 &    3 &  1& 0& 0\\
08379b & 25453 & 143  & 37.809$\;\;$ & 88.9 & 35.8 &  0.0 &   56 &  3& 0& 0\\
08379c$\dots$ & 26090 &   1  &   0.291 &  214.7 & $-$19.6 &    8.2 &  101 & 13& 0& 0\\
\hline
\end{tabular}
\end{center}
\setcounter{table}{2}
{ 
Table~2 is published in its entirety in the electronic edition 
of {\it The Astrophysical Journal}. See also 
ftp://ftp.astro.su.se/pub/head/grb/catalogs/etable2.txt.\\
A portion is shown here for guidance regarding its form and content.}\\
$^{\rm a}${Event identifier consisting of TJD  plus an identifying letter.}\\
$^{\rm b}${Start of the event within TJD.}\\
$^{\rm c}${BATSE trigger number or 1 for non-triggered events.}\\
$^{\rm d}${Peak flux in the 50 - 300 keV range.}\\
$^{\rm e}${The best fit coordinates (J2000) and the size of the 
$1\sigma$  confidence area.} \\
$^{\rm f}${Duration.} \\
$^{\rm g}${Number of 1.024 s bins in the original DISCLA data
where the signal exceeds 50\% of the peak value.}\\
$^{\rm h}${Flag identifying overlap with the K98 catalog. 
A 1 indicates existence in K98.}\\
$^{\rm i}${Flag characterizing the data quality with
non-zero values indicating proximity to the data gap.}\\
\label{generaltable}
\medskip
\normalsize

\footnotesize
\begin{center}
{\sc TABLE 3\\
The ``low brightness'' part of the efficiency matrix}  
\vskip 4pt
\begin{tabular}{rrrrrrrr} 
\hline
\hline
$\log c_m$ & \multicolumn{7}{c}{$\log c_e$} \\ 
\cline{2-8} \\
 &   $-1.3$  &  $-1.2$ & $-1.1$&  $-1.0$& $-0.9$& $-0.8$ & $-0.7$ \\
 \hline
 $-0.7$  & 0.0    & 0.0  & 0.0   & 0.0    & 0.7   & 11.7   &{\bf 43.1}\\ 
 $-0.8$  & 0.0    & 0.0  & 0.3   & 1.0    & 12.1&{\bf 36.5}& 11.0 \\ 
 $-0.9$  & 0.0    & 0.0  & 1.8   & 10.4 &{\bf 28.6}& 11.2  & 0.7  \\
 $-1.0$  & 0.0    & 0.8  & 6.8 &{\bf 15.9}& 9.5   & 1.3    & 0.4  \\
 $-1.1$  & 0.0    & 4.3&{\bf 6.9}& 6.5    & 1.0   & 0.1    & 0.0  \\
 $-1.2$  & 0.0 &{\bf 0.4}& 2.9   & 0.8    & 0.2   & 0.1    & 0.1  \\
 $-1.3$ &{\bf 0.0}& 1.2  & 0.0   & 0.0    & 0.0   & 0.0    & 0.0 \\
\hline 
\end{tabular}
\end{center}
\setcounter{table}{3}
{
Probability  $F(c_e,c_m)$  (in  \%)  for  the  detection  of a GRB  with an
expected  peak count,  $c_e$, and a measured  peak count rate,  $c_m$ 
($\counts\secinv\cminvsq$).  The
vertical  and  horizontal  bins are uniform in the  logarithm  of $c$.  The
lower edges of the bins are given in terms of $\log  (c_{e,m})$.   
}
\medskip
\normalsize
\end{document}